\documentclass[
prd,
reprint,
notitlepage,
nofootinbib,
nobibnotes,
amsmath,amssymb,
aps,
floatfix,
]{revtex4-2}

\usepackage{graphicx}
\usepackage{dcolumn}
\usepackage{bm}
\usepackage{amsmath,amssymb} 
\usepackage{amsfonts}
\usepackage{subfigure}
\usepackage{graphicx}
\usepackage{ulem}
\usepackage{color}
\usepackage{xcolor}
\usepackage{float}
\usepackage{makecell}
\usepackage{gensymb}
\usepackage{multirow}
\usepackage{ulem}
\usepackage{graphicx}
\usepackage{tabularx}
\usepackage{aas_macros}
\usepackage{url}
\usepackage{silence}
\usepackage{hyperref}
\usepackage[mathlines]{lineno}

\begin{document}

\preprint{APS/123-QED}

\title{Advancing weak lensing mass mapping with a mask-aware HEALPix transformer}

\author{Yihe Wang}
\affiliation{State Key Laboratory of Dark Matter Physics, School of Physics and Astronomy, Shanghai Jiao Tong University, Shanghai 200240, China}
\affiliation{Key Laboratory for Particle Astrophysics and Cosmology (MOE) / Shanghai Key Laboratory for Particle Physics and Cosmology, China}

\author{Yu Yu}
\email{yuyu22@sjtu.edu.cn}
\affiliation{State Key Laboratory of Dark Matter Physics, School of Physics and Astronomy, Shanghai Jiao Tong University, Shanghai 200240, China}
\affiliation{Key Laboratory for Particle Astrophysics and Cosmology (MOE) / Shanghai Key Laboratory for Particle Physics and Cosmology, China}

\received{11 November 2025}
\accepted{22 January 2026}
\published{26 February 2026}

\begin{abstract}
We present HEALFormer, a transformer-based neural network architecture for weak gravitational lensing mass mapping that reconstructs convergence maps from incomplete and noisy shear observations on the celestial sphere. 
The model operates directly on the Hierarchical Equal Area isoLatitude Pixelization and employs learnable mask tokens to handle arbitrary survey geometries without requiring preprocessing. Through a progressive training strategy, HEALFormer efficiently processes high-resolution maps up to $N_{\mathrm{side}}=1024$ and demonstrates excellent performance across diverse survey footprints including KiDS, DES, DECaLS, and Planck. The model generalizes robustly to cosmological parameters beyond its training set, producing nearly unbiased reconstructions with superior noise suppression compared to traditional Kaiser-Squires and Wiener filter methods. Remarkably, HEALFormer exceeds the theoretical phase recovery limits of linear reconstruction methods at small scales, achieving a fundamental breakthrough in weak lensing analysis. The combination of computational efficiency, reconstruction accuracy, and adaptability to varying survey configurations makes HEALFormer well-suited for current and next-generation cosmological surveys. Code is available at \href{https://github.com/lalalabox/healformers}{GitHub}.
\end{abstract}

\maketitle


\section{Introduction}\label{sec:intro}

Modern cosmological observations produce wide-field maps of the cosmic microwave background, galaxy angular distributions, and weak gravitational lensing effects that span substantial fractions of the celestial sphere. Traditional flat-sky approximations, which treat small patches as locally Euclidean, become inadequate as survey areas expand to such large angular scales. The Hierarchical Equal Area isoLatitude Pixelization (HEALPix) scheme \citep{GORSKI2005} has emerged as the standard framework for spherical data representation, providing equal-area pixelization and computational efficiency through fast spherical harmonic transforms. The next generation of wide-field missions will dramatically extend this paradigm. The Rubin Observatory Legacy Survey of Space and Time\footnote{\url{https://www.lsst.org/}} \citep{collaborationLargeSynopticSurvey2012}, Euclid\footnote{\url{https://www.euclid-ec.org/}} \citep{scaramellaEuclidPreparationEuclid2022}, and the China Space Station Survey Telescope\footnote{\url{https://www.bao.ac.cn/csst/}} \citep{yaoCSSTWLPreparation2023,collaborationIntroductionChinaSpace2025} will together map substantial portions of the sky at unprecedented depth. Extracting maximal cosmological information from these expanded surveys demands analysis methods that operate natively within the HEALPix framework while maintaining robustness against complex survey masks and spatially varying noise.

Weak gravitational lensing, the subtle distortion of galaxy shapes caused by matter along the line of sight, is a powerful probe of the cosmic mass distribution. It constrains fundamental cosmological parameters, including the matter fraction $\Omega_m$ and the clustering amplitude $\sigma_8$ (see review by, e.g., \citep{mandelbaumWeakLensingPrecision2018, Bartelmann2001}). The central task of mass mapping is to reconstruct the scalar convergence field $\kappa$ from the spin-2 shear $\gamma$ inferred from observed galaxy ellipticities. The scalar nature of $\kappa$ makes convergence maps more feasible to statistical analysis than the spin-2 shear field, establishing them as the preferred target for cosmological inference.

Mass mapping presents a challenging inverse problem due to two fundamental obstacles: shape noise and incomplete sky coverage. 
Shape noise arises because observed galaxy ellipticities combine intrinsic morphologies with the weak lensing shear signal, and this intrinsic scatter typically dominates the error budget in weak lensing analyses. Simultaneously, survey masks introduce spatial discontinuities that corrupt local and global mode recovery.
Classical methods address these challenges only partially. The Kaiser-Squires (KS) estimator \citep{kaiserMappingDarkMatter1993} provides a direct linear inversion but amplifies noise and introduces boundary artifacts. Wiener filtering (WF) \citep{lahavWienerReconstructionAllSky1994} regularizes the solution by imposing a Gaussian prior on the convergence power spectrum, at the cost of suppressing non-Gaussian features. The recently developed AKRA method \citep{shiAccurateKappaReconstruction2024, shiAKRA20Accurate2024} relaxes this assumption through a prior-free maximum likelihood formulation, though it still requires clean shear fields free of shape noise.
Sparsity-based methods move beyond the Gaussian framework. GLIMPSE~\citep{leonardGLIMPSEAccurate3D2014} and DarkMappy~\citep{priceSparseBayesianMass2021} employ tailored priors and hierarchical Bayesian formulations to recover non-Gaussian structures such as clusters and filaments. However, these approaches typically operate on flat-sky projections, decomposing the survey into overlapping patches that are reconstructed independently and stitched together. While this strategy captures small-scale features, it tends to suppress low-$\ell$ modes and remains sensitive to border effects. Hybrid methods such as MCALens~\citep{starckWeakLensingMass2021} combine linear and sparse components directly on the sphere, improving over standard WF, yet still rely on predefined priors. Consequently, most existing approaches either adopt flat-sky approximations, impose strong statistical assumptions, or struggle to accommodate complex mask geometries.

Deep learning improves the reconstruction on flat maps. DeepMass~\citep{jeffreyDeepLearningDark2020} leverages a U-Net architecture and achieves tighter cosmological constraints than KS, WF, and DarkMappy in Fisher forecasts \citep{grewalComparingMassMapping2024}. Generative adversarial networks denoise KS reconstructions and recover two statistics, though cross-correlation metrics can degrade \citep{shirasakiNoiseReductionWeak2021}. Probabilistic score-based models such as DLPosterior~\citep{remyProbabilisticMassMapping2023} perform competitively with WF in pixel space but underperform in power-spectrum ratios on $\kappa$TNG dataset. A common limitation of these works is their flat-sky formulation, which introduces projection artifacts and complicates the treatment of survey boundaries. Spherical learning seeks to avoid these issues by operating directly on HEALPix. Graph-based models such as DeepSphere~\citep{perraudinDeepSphereEfficientSpherical2019} demonstrate this direction by classifying two cosmological parameters from $\kappa$ on the sphere, but memory costs limit resolution and masks are typically handled by zero-filling, which degrades performance near boundaries. More recently, vision transformers (ViTs) have been adapted to HEALPix via sphere segmentation \citep{carlssonHEALSWINVisionTransformer2024}. Yet, scaling to high resolution and accommodating arbitrary mask geometries remain open challenges, underscoring the need for spherical-native, mask-aware learning.

We introduce HEALFormer (HF), a mask-aware transformer for weak lensing mass mapping on the sphere. The model consumes HEALPix maps directly and handles partial sky coverage and realistic noise natively. Inspired by masked autoencoders (MAE) \citep{heMaskedAutoencodersAre}, it employs an asymmetric encoder-decoder architecture that processes only visible and edge patches. This design reduces memory and computational demands while preserving low-$\ell$ information crucial for large-scale structure. We represent arbitrary mask geometries through learnable MASK tokens \citep{devlinBERTPretrainingDeep2019} that encode both masked pixels and fully masked patches, enabling principled inpainting and mitigating border artifacts. Our training strategy combines MAE retraining with Low-Rank Adaptation (LoRA) \citep{huLoRALowRankAdaptation2021} for efficient fine-tuning. We adopt a progressive coarse-to-fine schedule by first training at $N_{\rm side}=256$ and then scaling to $N_{\rm side}=1024$ on standard hardware. Evaluations across DES \citep{jeffreyDarkEnergySurvey2021}, KiDS \citep{heymansKiDS1000CosmologyMultiprobe2021}, DECaLS \citep{deyOverviewDESILegacy2019}, and Planck \citep{planckcollaborationPlanck2018Results2018} footprints demonstrate improved reconstruction over classical methods in both pixel space and harmonic space, indicating a scalable path for forthcoming surveys.

The paper is organized as follows. Section~\ref{sec:theory} reviews weak lensing theory on the sphere and classical reconstruction methods. Section~\ref{sec:deeplearning} presents the \texttt{HEALFormer} architecture and implementation. Section~\ref{sec:result} compares performance against traditional techniques using map-level and spectral metrics. We conclude with a summary and outlook in Sec.~\ref{sec:conclusion}.

\section{Weak gravitational lensing on the sphere}\label{sec:theory}

The theoretical foundation of weak lensing begins with the relationship between the gravitational potential $\Phi$ and the matter overdensity field $\delta \equiv \delta\rho / \bar{\rho}$. These scalar fields are connected through the Poisson equation:

\begin{equation}
  \label{eq:poisson}
  \nabla^2_r \Phi(t, \boldsymbol{r}) = \frac{3 \Omega_m H_0^2}{2 a(t)} \delta(t, \boldsymbol{r}) \ ,
\end{equation}

\noindent where $\nabla^2_r$ is the spatial Laplacian, $\Omega_m$ is the matter density parameter, $H_0$ is the Hubble constant, and $a(t) = 1/(1+z)$ is the scale factor at redshift $z$.

The lensing potential $\phi$ represents the integrated effect of gravitational fields along the line of sight. Using spherical coordinates $(\chi, \theta, \varphi)$ where $\chi$ is the comoving radial distance and $(\theta, \varphi)$ are angular coordinates on the celestial sphere, the lensing potential is obtained through:

\begin{equation}
  \label{eq:born}
  \phi(\chi,\theta,\varphi) = \frac{2}{c^2} \int_0^{\chi} d\chi' \frac{f_K(\chi-\chi')}{f_K(\chi)f_K(\chi')} \Phi(\chi',\theta,\varphi).
\end{equation}
Here, $f_K$ denotes the angular diameter distance function that depends on the spatial curvature $K$. For closed, flat, or open universes, $f_K$ takes the form of $\sin$, identity, or $\sinh$, respectively. This expression employs the Born approximation, treating light paths as unperturbed by gravitational deflections.
Given a population of source galaxies, we obtain the two-dimensional lensing potential on the celestial sphere by integrating over the source galaxy distribution $n(z)$ as

\begin{equation}\label{eq:lensing_potential}
  \phi(\theta,\varphi) = \int d\chi \ n(z(\chi)) \ \phi(\chi,\theta,\varphi),
\end{equation}

\noindent yielding a scalar field.

The mathematical framework for spherical lensing employs spin-weighted spherical harmonics $_s Y_{lm}(\theta,\varphi)$ with spin-weight $s$ \citep{Castro-PhysRevD.72.023516}. The covariant derivatives $\eth$ and $\bar{\eth}$ act as raising and lowering operators for the spin weight:

\begin{equation}
  \eth _s Y_{lm} = -\sqrt{(l-s)(l+s+1)} \ _{s+1} Y_{lm},
\end{equation}

\begin{equation}
  \bar{\eth} _s Y_{lm} = \sqrt{(l+s)(l-s+1)} \ _{s-1} Y_{lm}.
\end{equation}
The lensing potential $\phi$ determines two fundamental observables, namely, the convergence $\kappa = \kappa_E + i\kappa_B$ (spin-0) and the shear $\gamma = \gamma_1 + i\gamma_2$ (spin-2). These quantities relate to the lensing potential through differential operators as follows:
\begin{equation}
  \label{eq:kappa}
    \kappa = -\frac{1}{2} \nabla^2 \phi ,
\end{equation}

\begin{equation}
  \label{eq:gamma}
    \gamma = -\frac{1}{4} (\eth^2 + \bar{\eth}^2)  \phi ,
\end{equation}

\noindent where $\nabla^2 = (\eth \bar{\eth} +\bar{\eth} \eth)$ is the spherical Laplacian operator.

The fundamental relationship between shear and convergence emerges naturally in harmonic space through their spherical harmonic coefficients. For the shear field:

\begin{equation}
  \label{eq:gammaSum}
  \gamma = \sum_{\ell m} \hat{\gamma}_{\ell m} \, _2Y_{\ell m},
\end{equation}

\noindent where the coefficients are obtained through

\begin{equation}
  \label{eq:gammaInt}
  \hat{\gamma}_{\ell m}  = \int d\Omega \ \gamma(\theta ,\varphi) \, _2Y_{\ell m}^{*}(\theta ,\varphi).
\end{equation}
The harmonic coefficients can be decomposed into $E$-mode and $B$-mode components as $\hat{\kappa}_{\ell m} = \hat{\kappa}_{E,\ell m} + i \hat{\kappa}_{B,\ell m}$ and $\hat{\gamma}_{\ell m} = \hat{\gamma}_{E,\ell m} + i \hat{\gamma}_{B,\ell m}$. In the harmonic domain, the relationships between the lensing potential, convergence, and shear take the following elegant form:
\begin{equation}
\hat{\kappa}_{\ell m}  = - \frac{1}{2} \ell (\ell+1) \hat{\phi}_{\ell m}
\end{equation}

\noindent and

\begin{equation}
\hat{\gamma}_{lm} = \frac{1}{2}\sqrt{(\ell-1)\ell(\ell+1)(\ell+2)}\hat{\phi}_{\ell m},
\end{equation}
where $\hat{\phi}_{\ell m}$ is the spherical harmonic transform of the lensing potential in Eq.~(\ref{eq:lensing_potential}). Combining these equations, we derive the relation:

\begin{equation}
\label{eq:k2g_harmonic}
\hat{\gamma}_{lm} = -\sqrt{\frac{(\ell-1)(\ell+2)}{\ell(\ell+1)}} \hat{\kappa}_{\ell m}.
\end{equation}

\noindent This fundamental equation provides the theoretical basis for all mass mapping techniques by directly linking observable shear to the desired convergence field.

The practical implementation of mass mapping requires expressing the shear-convergence relationship in discrete form. Using matrix notation, the observation equation becomes

\begin{equation} \label{eq:g2k-matrix}
  \boldsymbol{\gamma} = \mathbf{A} \boldsymbol{\kappa} + \mathbf{n} \ ,
\end{equation}

\noindent where $\boldsymbol{\gamma}$ represents the observed shear field discretized into pixels, $\mathbf{A}$ encodes the lensing transformation \citep{Bartelmann2001}, and $\mathbf{n}$ represents observational noise.

\subsection{Kaiser and Squires}

With the harmonic relationship specified, we briefly review classical estimators, beginning with the direct inversion of KS~\citep{kaiserMappingDarkMatter1993}. It represents the most direct approach to mass reconstruction, exploiting the linear relationship between shear and convergence in harmonic space. This technique provides an exact solution under ideal conditions but neglects observational complications including noise and incomplete sky coverage.

The KS reconstruction directly inverts Eq.~(\ref{eq:k2g_harmonic}) to recover convergence from observed shear:

\begin{equation}
  \label{eq:g2k_harmonic}
  \hat{\kappa}_{\ell m} = \hat{\kappa}_{E, \ell m} + i\hat{\kappa}_{B, \ell m} = -\sqrt{\frac{\ell(\ell+1)}{(\ell-1)(\ell+2)}} \hat{\gamma}_{\ell m}.
\end{equation}
A fundamental limitation of this approach is the mass-sheet degeneracy \citep{bartelmannClusterMassEstimates1994}, which manifests as an undefined $\ell=0$ mode. This degeneracy prevents determination of the mean convergence from shear measurements alone. Following standard practice for large-scale surveys \citep{masseyDarkMatterMaps2007}, we assume zero mean convergence by setting the monopole to zero.

As one of the methods in comparison, our KS implementation utilizes healpy spherical harmonic transforms with the maximum multipole restricted to $\ell_{\max}=2N_{\rm side}$ to ensure numerical stability \citep{jeffreyDarkEnergySurvey2021}, providing sufficient resolution while maintaining computational tractability.

\subsection{Wiener filter}

The Wiener filter provides an optimal linear reconstruction under Gaussian assumptions by balancing data fidelity with prior knowledge of signal and noise statistics. This Bayesian approach seeks the convergence field $\boldsymbol{\kappa}$ that maximizes the posterior probability given observed shear $\boldsymbol{\gamma}$.

The method requires specification of the signal covariance $\mathbf{S}_\kappa=\langle \boldsymbol{\kappa} \boldsymbol{\kappa}^\dagger \rangle$ and noise covariance $\mathbf{N}=\langle \mathbf{n} \mathbf{n}^\dagger \rangle$ \citep{wiener1949extrapolation, zaroubiWienerReconstructionLargescale1995, jeffreyFastSamplingWiener2018}. For isotropic convergence fields, $\mathbf{S}_\kappa$ becomes diagonal in harmonic space with elements given by the power spectrum $C_\kappa(\ell)$. Similarly, uncorrelated Gaussian shape noise yields a diagonal noise covariance in pixel space.

Under Gaussian assumptions for both signal and noise, the posterior probability maximization yields the Wiener filter solution through minimization of the $\chi^2$ statistic:

\begin{equation}
  \chi^2 = {(\boldsymbol{\gamma} - \mathbf{A} \boldsymbol{\kappa})}^\dagger \mathbf{N}^{-1} (\boldsymbol{\gamma} - \mathbf{A} \boldsymbol{\kappa}) + \boldsymbol{\kappa}^\dagger \mathbf{S}_\kappa^{-1} \boldsymbol{\kappa}.
\end{equation}

\noindent This optimization balances data fidelity (first term) with prior expectations (second term), yielding the optimal estimate:

\begin{equation}\label{eq:wf_solution}
  \boldsymbol{\kappa}_{\rm WF} = {(\mathbf{S}_{\kappa}^{-1} + \mathbf{A}^\dagger \mathbf{N}^{-1} \mathbf{A})}^{-1} \mathbf{A}^\dagger \mathbf{N}^{-1} \boldsymbol{\gamma}.
\end{equation}

\noindent This solution represents the optimal linear estimator under Gaussian assumptions, providing the minimum variance unbiased reconstruction of the convergence field.

Direct implementation faces computational challenges for high-resolution maps due to large matrix inversions. We employ the DANTE algorithm \citep{ramanahWienerFilteringPure2019}, which exploits the diagonal structure of covariance matrices in their natural bases (pixel space for noise, harmonic space for signal) to enable iterative solutions without explicit matrix inversions. This approach has proven practical for large-scale surveys, as demonstrated in DES-Y3 analysis \citep{jeffreyDarkEnergySurvey2021}.

\section{Deep learning mass mapping}\label{sec:deeplearning}

Deep learning has revolutionized numerous scientific domains, with transformer architectures achieving exceptional performance in natural language processing and computer vision. This section describes our adaptation of transformer models to the spherical mass mapping problem.

\subsection{Vision transformers}\label{sec:vit}

Transformers have become a standard tool across machine learning because they model long-range interactions efficiently. Their core operation is self-attention \citep{vaswaniAttentionAllYou2023}, which forms data-dependent weighted averages of inputs:
\begin{equation}\label{eq:attention}
\operatorname{Attention}(Q, K, V)=\operatorname{softmax}\!\left(\frac{Q K^{T}}{\sqrt{d}}\right) V,
\end{equation}
where $Q$, $K$, and $V$ are the query, key, and value matrices obtained from learnable projections of the input, and $d$ is the key dimension. Unlike CNNs that emphasize local neighborhoods, attention connects all positions at once, capturing global structure and accommodating variable input sizes. ViTs adapt this idea to images by splitting an image into fixed-size patches and treating the resulting sequence of patch embeddings as tokens \citep{dosovitskiyImageWorth16x162020}. Patch size controls a simple trade-off: smaller patches offer finer detail but increase the token count, and the cost of self-attention scales quadratically with the number of tokens [$\mathcal{O}(N^2)$ for $N$ patches]. Following common practice, we use $16\times16$ patches to balance resolution and efficiency. In our spherical setting, we apply the same patchification to HEALPix maps, arranging patches using the NEST ordering so that nearby pixels map to nearby tokens in the sequence.

MAEs further strengthen this framework through self-supervised pretraining \citep{heMaskedAutoencodersAre}. An asymmetric encoder processes only the visible patches while a lightweight decoder reconstructs the masked ones. This forces the model to learn robust representations of structure and context without labeled targets, and these representations transfer effectively to downstream tasks. Although natural images and cosmological fields differ, the priors learned by MAE---patterns, continuity, and multiscale context---are useful for weak lensing mass mapping. In practice, MAE pretraining accelerates convergence, improves denoising and inpainting in masked regions, and reduces the amount of task-specific training required. With these ingredients, we now specify a spherical-native architecture tailored to HEALPix data.

\subsection{HEALFormer architecture}\label{sec:HEALFormer}

\begin{figure*}
  \centering
  \includegraphics[width=0.9\textwidth]{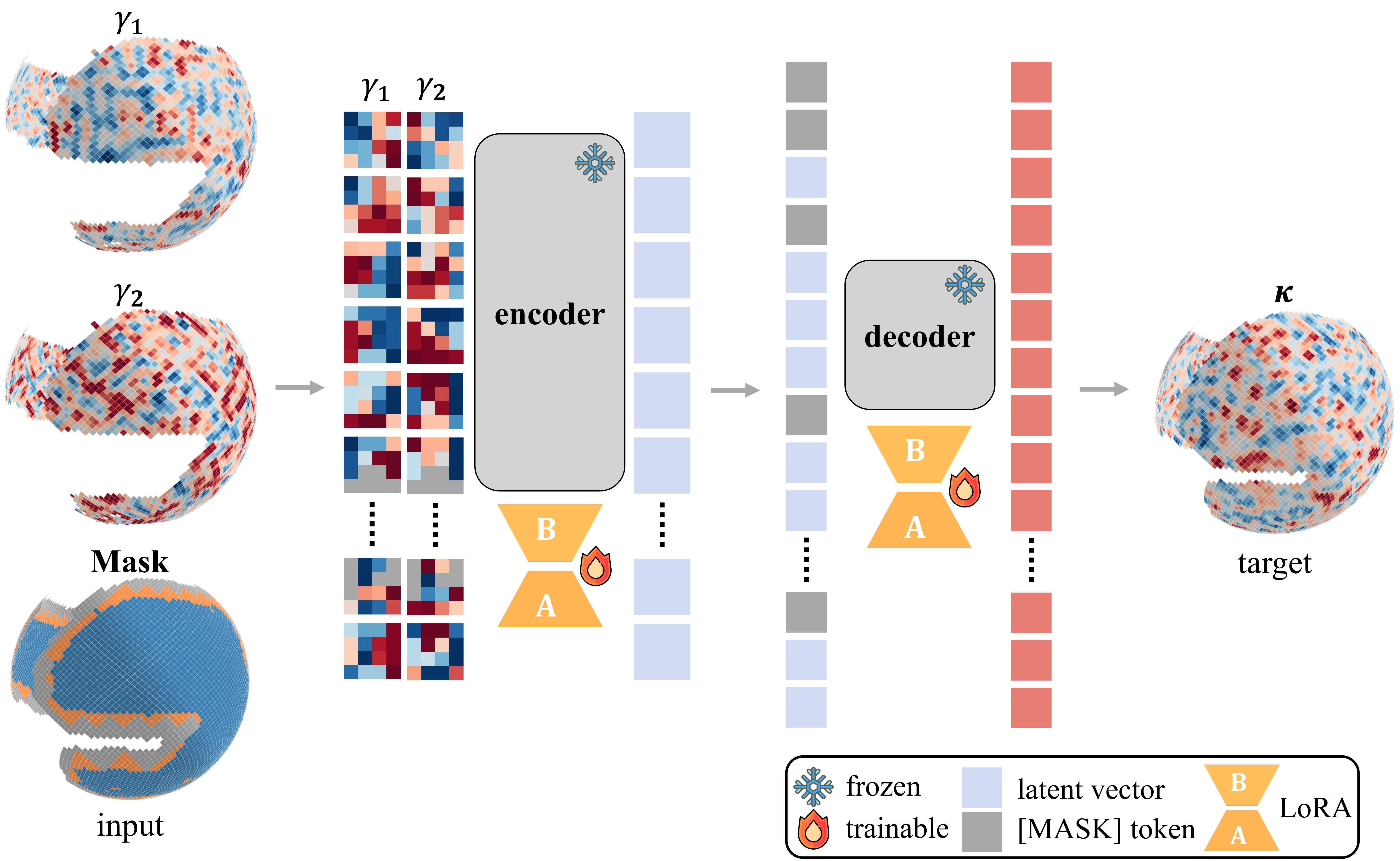}
  \caption{
  HEALFormer architecture. The model reconstructs convergence maps ($\kappa$) from masked, noisy HEALPix shear inputs ($\gamma_1,\gamma_2$) and an explicit mask. Pixels are classified as visible (blue), inner edge (orange), or outer edge (grey). The visible and inner-edge patches will be projected into patch embddding (inserting the MASK token for the orange inner-edge pixels). The asymmetric encoder\mbox{-}decoder processes these embeddings through the encoder, then inserts learnable MASK tokens for outer-edge patches before decoding. This design enables joint denoising and inpainting, yielding a complete $\kappa$ field over both observed and masked regions. Efficient fine-tuning via LoRA updates only $\sim$10\% of parameters while the backbone remains frozen.}\label{fig:architecture}
\end{figure*}

HEALFormer (Fig.~\ref{fig:architecture}) is a transformer tailored for spherical mass mapping.
The model takes as input the shear components $\gamma_1$, $\gamma_2$, and a binary survey mask. Because weak lensing inversion is nonlocal, reconstructing convergence near survey boundaries requires information from an extended region. We therefore define an edge zone around the survey footprint (see Sec.~\ref{sec:data_generation} for details) and partition the sky into three mutually exclusive patch categories. A patch is \textit{visible} if all its pixels are observed and none lie in the edge zone, \textit{inner edge} if it contains a mixture of observed and edge pixels, and \textit{outer edge} if it lies entirely within the edge zone. This classification concentrates learning on boundary neighborhoods while preserving clean supervision in the survey interior.

The forward pass proceeds as follows. Input maps are partitioned into $16\times16$ patches under HEALPix NEST ordering, which naturally groups spatially adjacent pixels. Each patch is projected into a $d_{\mathrm{model}}$-dimensional embedding; for inner-edge patches, learnable MASK tokens replace masked pixel positions within this embedding. Positional embeddings encoding HEALPix geometry are then added. The encoder, a stack of transformer blocks with self-attention and feed-forward layers, processes only visible and inner-edge embeddings. This asymmetric design improves computational efficiency by excluding outer-edge patches from the expensive encoding stage. After encoding, MASK tokens for outer-edge patches are inserted into the latent sequence and positional embeddings are reapplied. The decoder predicts convergence values from this augmented representation. The backbone follows the MAE paradigm and remains frozen during training; only LoRA parameters are updated (Sec.~\ref{sec:training}), reducing trainable parameters to roughly 10\%.

This architecture addresses three practical challenges. First, \textit{denoising}: shape noise dominates at high resolution, so features must be extracted without suppressing cosmological signal. Second, \textit{inpainting}: the nonlocal shear-to-convergence relation permits recovery in masked regions from surrounding context. Third, \textit{border effects}: complex survey geometries demand explicit mask-aware handling. The asymmetric encoder decoder tackles these jointly: the encoder denoises and compresses information from observed patches; the decoder, guided by MASK tokens, predicts a seamless $\kappa$ field that extends into masked borders. Self-attention propagates information across the mask topology, improving boundary fidelity.

Three design choices underpin this approach. Learnable mask representation: rather than zero-filling, MASK tokens explicitly encode missing data, allowing the model to learn mask geometry. Asymmetric processing: restricting the encoder to visible and inner-edge patches reduces computation on partial-sky data while the decoder reconstructs borders from encoded context. Universal mask handling: a single model accommodates diverse survey footprints via variable-length sequences and specialized position embeddings (Appendix~\ref{app:position_embedding}) that respect spherical geometry.

\subsection{Mask token strategy}\label{sec:mask_token}

Astronomical observations inevitably contain masked regions from survey boundaries, bright-star exclusions, low signal-to-noise areas, and Galactic contamination. Traditional methods address these through preprocessing: KS+ \citep{piresEuclidReconstructionWeaklensing2020} applies discrete cosine transform inpainting, while DeepMass~\citep{jeffreyDeepLearningDark2020} uses zero-filling. HEALFormer instead employs learnable MASK tokens, inspired by their success in language models \citep{devlinBERTPretrainingDeep2019}. These trainable vectors encode the statistical properties and spatial context of masked regions.

Because maps are divided into patches, MASK tokens operate at two levels. At the \textit{pixel level}, inner-edge patches contain both observed and masked pixels; MASK tokens replace masked positions within the patch embedding before encoding, enabling the encoder to distinguish observations from gaps. At the \textit{patch level}, MASK tokens for outer-edge patches are appended to the latent sequence after encoding but before decoding, enabling the decoder to predict $\kappa$ values beyond survey boundaries from visible-region context.

A single learnable MASK embedding is shared across all positions and maps; spatial information is injected through additive positional embeddings. This explicit representation allows HEALFormer to capture the geometry of survey masks and mitigate boundary artifacts where traditional methods struggle.

\subsection{Data generation and preprocessing}\label{sec:data_generation}

We follow the DES-Y3 setup \citep{jeffreyDarkEnergySurvey2021} to simulate weak lensing data. Cosmological parameters are sampled from the KiDS-450 posterior \citep{Hildebrandt2017}. Using \texttt{PYCCL} \citep{chisariCoreCosmologyLibrary2019}, we compute the theoretical lensing power spectra, then use \texttt{HEALPY} to draw Gaussian full-sky realizations of shear ($\gamma$) and convergence ($\kappa$) on the HEALPix grid. Maps are generated at two resolutions, $N_{\rm side}=256$ and $N_{\rm side}=1024$. For efficiency, we create 10{,}000 realizations at $N_{\rm side}=256$ and 3{,}000 at $N_{\rm side}=1024$, and split each set $90\%/10\%$ into training and testing.

To evaluate performance under realistic sky coverage, we apply survey masks from KiDS, DES, DECaLS, and Planck. For masks defined at higher resolution (for example Planck at $N_{\rm side}=2048$), we change resolution in harmonic space by discarding modes above $\ell_{\max}$. This avoids aliasing when degrading to $N_{\rm side}=256$ or $1024$. We also build an explicit boundary zone around each footprint by iteratively expanding the mask. Starting from the original mask, we call healpy.get\_all\_neighbours 16 times (consistent with the patch size) to match the patch size and mark the resulting 16 pixel thick ring as the edge pixels. This boundary is important because the shear to convergence relation is nonlocal, and information near the survey edge helps reconstruct the convergence on both sides of the mask.

For consistent comparison across experiments, we simulate isotropic Gaussian shape noise with a galaxy number density of $n_{\rm gal}=30$ galaxies/arcmin$^2$ (representative of Euclid \citep{scaramellaEuclidPreparationEuclid2022} and LSST-Y10 \citep{changEffectiveNumberDensity2013}).
In order to make sure the noise is independent for each sky realization, we generate the noise on-the-fly during training and testing rather than stored in the dataset, for each sky realization, we use a new random seed to generate the noise.

The standard deviation of the noise $n$ in Eq.~(\ref{eq:g2k-matrix}) is given by:
\begin{equation}
  \sigma_n = \frac{\sigma_\epsilon}{\sqrt{N_g}},
\end{equation}

\noindent where $\sigma_\epsilon$ represents the intrinsic shape dispersion combined with measurement uncertainty, and $N_g$ is the number of galaxies per pixel. Following common practice \citep{schrabbackPreciseWeakLensing2018,jeffreyDeepLearningDark2020,piresEuclidReconstructionWeaklensing2020, fiedorowiczKaRMMaKappaReconstruction2022a}, we adopt $\sigma_\epsilon=0.4$. For individual ellipticity components, this translates to $\sigma_{\epsilon_1}=\sigma_{\epsilon_2}=\sigma_\epsilon/\sqrt{2}\approx 0.28$, with $\sigma_\epsilon=\sqrt{\sigma_{\epsilon_1}^2+\sigma_{\epsilon_2}^2}$. We calculate $N_g$ for each pixel using the nside2pixarea function from \texttt{HEALPY} to determine pixel area, then multiplying by the galaxy number density. 

\subsection{Training}\label{sec:training}

\subsubsection{Fine-tuning}

We initialize HEALFormer from a pretrained MAE (ViT-Base/16) to transfer generic image priors to the weak lensing setting. This initialization stabilizes optimization and substantially reduces both training time and compute relative to training from scratch. To adapt the backbone efficiently, we employ LoRA \citep{huLoRALowRankAdaptation2021}, a parameter-efficient fine-tuning technique. Rather than retraining all network weights, LoRA freezes the pretrained parameters and introduces a small set of trainable modifications that capture task-specific adjustments. This strategy achieves performance comparable to full retraining at a fraction of the memory and computation cost \citep{shuttleworthLoRAVsFull2024}, while also allowing multiple survey configurations to be stored compactly as lightweight parameter sets rather than duplicating the entire model.

In our configuration, we train the query, key, and value projections of each attention block, the normalization layers, the learnable MASK token, and the position embeddings; all remaining parameters are frozen. The full model contains $\sim10^8$ parameters, of which $\sim10^7$ ($\approx10\%$) are trainable under LoRA.\@ This setup preserves capacity where it matters for adaptation while keeping the training and storage footprint modest, enabling us to fine-tune at the resolutions used in this work.

\subsubsection{Optimizer and learning rate strategy}
The learning rate is a critical hyperparameter in deep learning that controls the step size during gradient descent optimization. Conventional approaches typically employ a cosine decay learning rate scheduler, which gradually reduces the learning rate throughout training to ensure smaller optimization steps as the model approaches convergence. 

For our model, we adopt the Schedule-Free AdamW optimizer recently proposed by \citet{defazioRoadLessScheduled2024}. This advanced optimizer eliminates the need for manual learning rate scheduling while maintaining the benefits of the original AdamW algorithm~\citep{kingma2014adam}. A key advantage of Schedule-Free AdamW is its robustness to a wider range of initial learning rates, reducing the need for extensive hyperparameter tuning.

In our implementation, we initialize the base learning rate at $1 \times 10^{-4}$. This approach significantly improves both convergence speed and final model performance compared to traditional schedulers, allowing us to achieve optimal results with less computational overhead and manual intervention.

\subsubsection{Coarse-to-fine training strategy}\label{sec:coarse-to-fine}
Training transformer-based models on high-resolution data (e.g., $N_{\mathrm{side}}=1024$) presents significant computational challenges due to the quadratic complexity $\mathcal{O}(N^2)$ of attention mechanisms. When comparing $N_{\mathrm{side}}=1024$ to $N_{\mathrm{side}}=256$, the number of pixels increases by a factor of 16, resulting in a sequence length that is 16 times longer. Consequently, the computational requirements increase by a factor of $16^2=256$ due to the transformer's quadratic complexity. This dramatic increase leads to substantially longer training times and significantly higher GPU memory consumption.

To effectively manage computational constraints while ensuring robust model performance, we adopt a progressive, two-stage coarse-to-fine training strategy. This method aligns with standard machine learning practices by gradually increasing the complexity of the task.

In the initial stage, the model is trained at a lower resolution ($N_{\mathrm{side}}=256$). This phase is computationally less demanding and presents a simplified mass-mapping scenario with smaller shape noise per pixel. Here, the model learns the essential relationship between shear ($\gamma$) and convergence ($\kappa$), acquiring fundamental skills for managing mask effects and performing basic denoising.

The second phase trains at higher resolution ($N_{\mathrm{side}}=1024$), where per-pixel noise increases substantially. This stage refines the model's denoising capability, emphasizing the extraction of meaningful signal from noise-dominated data. The progressive increase in task complexity allows the model to develop the advanced representations required for accurate high-resolution reconstruction.

Throughout both stages we retain the $16\times16$ patch size. Although smaller patches can improve performance \citep{nguyenImageWorthMore2025}, reducing patch size to $8\times8$ quadruples the sequence length, exceeding single-GPU memory at $N_{\mathrm{side}}=1024$. Maintaining a fixed patch size also simplifies coarse-to-fine transfer: the pretrained low-resolution weights load directly into the high-resolution model, enabling parameter-efficient fine-tuning with LoRA rather than full retraining.

\subsubsection{Loss function}\label{sec:loss_function}
We train the model with a hierarchical pixel-space loss that aggregates mean squared error (MSE) terms evaluated at multiple spatial resolutions. This multiscale design guides the network to capture large-scale structure while preserving small-scale detail, which improves convergence-map reconstruction. We explored adding harmonic-space losses, but found no measurable gain over pixel-space losses while incurring substantially higher computational cost.

Balancing multiple loss terms is nontrivial. We follow the uncertainty-based weighting of \citet{cipollaMultitaskLearningUsing2018}, which treats each term as having a task-dependent, homoscedastic uncertainty. The associated weights are learned jointly with the model, removing manual tuning and allowing losses with different units and dynamic ranges to be optimized together.

Formally, the total loss comprises multiscale terms evaluated over visible pixels plus two additional terms that target boundary regions:
\begin{align}\label{eq:loss}
  \mathcal{L} &= \sum_{i=1}^{N_{\mathrm{scale}}} \biggl( \frac{1}{\omega_i^2}\,\mathcal{L}_i + \log \omega_i \biggr) \notag\\
  &\quad + \sum_{k \in \{\mathrm{inner},\,\mathrm{outer}\}} \biggl( \frac{1}{\omega_k^2}\,\mathcal{L}_k + \log \omega_k \biggr),\\
  \mathcal{L}_i &= \frac{1}{P_i}\sum_{j=1}^{P_i}\bigl[\,g_i(\kappa^{\mathrm{pred}})_j - g_i(\kappa^{\mathrm{true}})_j\,\bigr]^2\ ,
\end{align}
where $P_i$ is the number of visible pixels at the $i$th resolution. The operator $g_i(\cdot)$ downsamples a map to resolution $N_{\mathrm{side}}/2^{i-1}$ via average pooling, so that the native resolution corresponds to $i=1$. Each learnable weight $\omega_i > 0$ (initialized to unity) encodes the uncertainty of its associated term; the $\log\omega_i$ regularizer prevents the solution from trivially driving any weight to zero. The boundary terms $\mathcal{L}_{\mathrm{inner}}$ and $\mathcal{L}_{\mathrm{outer}}$ are computed at the native resolution but restricted to inner-edge and outer-edge pixels, respectively, which lie immediately adjacent to the mask and are more difficult to reconstruct than interior visible pixels.

We choose $N_{\mathrm{scale}} = \log_2 N_{\mathrm{side}} + 1$ scales corresponding to $N_{\mathrm{side}} \in \{1, 2, 4, \ldots, N_{\mathrm{side}}\}$. For $N_{\mathrm{side}}=256$ this yields 9 scale terms; together with the two boundary terms the loss sums 11 uncertainty-weighted components. When computing MSE at any scale, we restrict the evaluation to visible and edge pixels and exclude masked areas that lie far from the survey boundary.

\subsection{Evaluation metrics}

We assess reconstruction fidelity in pixel space and harmonic space. Pixel-space metrics test pointwise agreement and distributional similarity, while harmonic-space metrics probe scale-dependent consistency.

\subsubsection{Pixel-space metrics}

We quantify multiplicative bias by fitting a constrained linear model between the reconstructed and true convergence maps,
\begin{equation}
  \kappa_{\mathrm{rec}} = m\,\kappa_{\mathrm{true}},
\end{equation}
after subtracting the mean of each map and restricting to unmasked pixels. Fixing the intercept to zero reflects the vanishing mean of the convergence field and isolates a single scale parameter $m$.

To obtain a slope estimate that is symmetric under exchanging the roles of the two maps, we minimize the average squared orthogonal distance to the line in the $(\kappa_{\mathrm{true}},\kappa_{\mathrm{rec}})$ plane,
\begin{equation}
  \mathcal{L}_{\mathrm{slope}}=\frac{1}{N}\sum_{i=1}^{N}{\left(\frac{m\,\kappa^{\mathrm{true}}_i-\kappa^{\mathrm{rec}}_i}{\sqrt{m^2+1}}\right)}^2.
\end{equation}
This total least-squares formulation ensures permutation symmetry: fitting $\kappa_{\mathrm{rec}}=m_1\,\kappa_{\mathrm{true}}$ and $\kappa_{\mathrm{true}}=m_2\,\kappa_{\mathrm{rec}}$ yields $m_1 m_2=1$, so the result does not depend on which map is taken as the predictor.

Overall amplitude-agnostic accuracy is reported using the normalized root mean squared error (NRMSE)~\citep{stephenImprovedNormalizationTimelapse2014},
\begin{equation}
  \mathrm{NRMSE}(\kappa_{\mathrm{rec}},\kappa_{\mathrm{true}})
  = \frac{\mathrm{RMSE}(\kappa_{\mathrm{rec}},\kappa_{\mathrm{true}})}{\sigma_{\kappa_{\mathrm{true}}}},
\end{equation}
with $\mathrm{RMSE}(x,y)=\sqrt{N^{-1}\sum_{i=1}^{N}{(x_i-y_i)}^2}$ and $\sigma_{\kappa_{\mathrm{true}}}$ the standard deviation of the true map. This normalization yields a dimensionless quantity comparable across data sets with different signal levels.

Linear association is summarized by the Pearson correlation coefficient,
\begin{equation}
  \rho(X,Y)=\frac{\mathrm{Cov}(X,Y)}{\sigma_X \sigma_Y}
  =\frac{\frac{1}{N}\sum_{i=1}^{N}(X_i-\bar{X})(Y_i-\bar{Y})}{\sigma_X \sigma_Y},
\end{equation}
which lies in $[-1,1]$ and is invariant to swapping $X$ and $Y$. Here $X$ and $Y$ denote the pixel values of the reconstructed and true maps on the unmasked domain.

To compare one-point distributions, we use the Jensen–Shannon (JS) divergence, as in the DES-Y1 mass-mapping analysis~\citep{mawdsleyDarkEnergySurvey2020},
\begin{equation}
  D_{\mathrm{JS}}(A,B)=\tfrac{1}{2}\!\left[D_{\mathrm{KL}}(A\Vert M)+D_{\mathrm{KL}}(B\Vert M)\right],
\end{equation}
where $M=(A+B)/2$, and with Kullback-Leibler divergence
\begin{equation}
  D_{\mathrm{KL}}(A\Vert B)=\sum_x A(x)\,\log_2\!\frac{A(x)}{B(x)}.
\end{equation}
Using base-2 logarithms, $D_{\mathrm{JS}}\in[0,1]$, where 0 indicates identical histograms and 1 indicates disjoint support. This complements pointwise metrics by testing distributional agreement.

\subsubsection{Harmonic-space metrics}

We test scale-dependent consistency through the $E$-mode power spectra of the reconstructed convergence,
\begin{align}
  C_{\ell}^{EE} &= \frac{1}{2\ell+1}\sum_{m}\left|\hat{\kappa}_{E,\ell m}\right|^2.
\end{align}
Spectra are computed with \texttt{HEALPY}'s \texttt{ANAFAST} using the same sky mask for all methods to ensure a fair comparison. While pseudo-$C_\ell$ estimators such as NaMaster~\citep{Alonso_2019} can correct for mask coupling, our consistent pipeline suffices for relative benchmarking. In the absence of systematics on the full sky, $C_{\ell}^{BB}=0$ for lensing convergence; any detected B-modes thus trace noise, masking, or residual systematics.

Scale-by-scale phase agreement with the truth is measured by the cross-correlation coefficient
\begin{equation}\label{eq:coefficient}
  r_{\ell}=\frac{C_{\ell}^{\mathrm{rec},\,\mathrm{true}}}{\sqrt{C_{\ell}^{\mathrm{rec}}\,C_{\ell}^{\mathrm{true}}}},
\end{equation}
which lies in $[-1,1]$. Values near unity indicate high phase coherence and accurate recovery at multipole $\ell$.

\section{Results}\label{sec:result}

We evaluate HEALFormer performance through comprehensive testing across various observational conditions, comparing against traditional KS and WF methods. Unless specified otherwise, all evaluations use cosmological parameters randomly drawn from the KiDS-450 posterior distribution \citep{Hildebrandt2017}.

\subsection{Mask effects}
To isolate the impact of incomplete sky coverage from observational noise, we first evaluate reconstruction performance on noise-free shear fields. We train and test HF at $N_{\mathrm{side}}=256$ using clean shear inputs with realistic survey masks applied. Typical masks comprise large contiguous footprints punctuated by small holes due to bright stars or low signal-to-noise regions.

Under these idealized conditions, WF achieves near-perfect reconstruction in both pixel and harmonic space, consistent with the analytic results of AKRA~\citep{shiAccurateKappaReconstruction2024,shiAKRA20Accurate2024}. KS degrades relative to full-sky performance, exhibiting increased scatter due to boundary artifacts. HF matches WF accuracy across all metrics (see Appendix~\ref{app:mask_effect} for quantitative details). These results confirm that, for typical survey geometries with largely contiguous coverage, the observed shear field retains sufficient information to recover the convergence exactly. Consequently, the absence of noise removes the primary challenge, and more sophisticated methods offer no additional benefit in this limit.

\subsection{Noise effect}\label{sec:noise_effect}

\begin{figure*}
  \centering
  \includegraphics[width=0.9\textwidth]{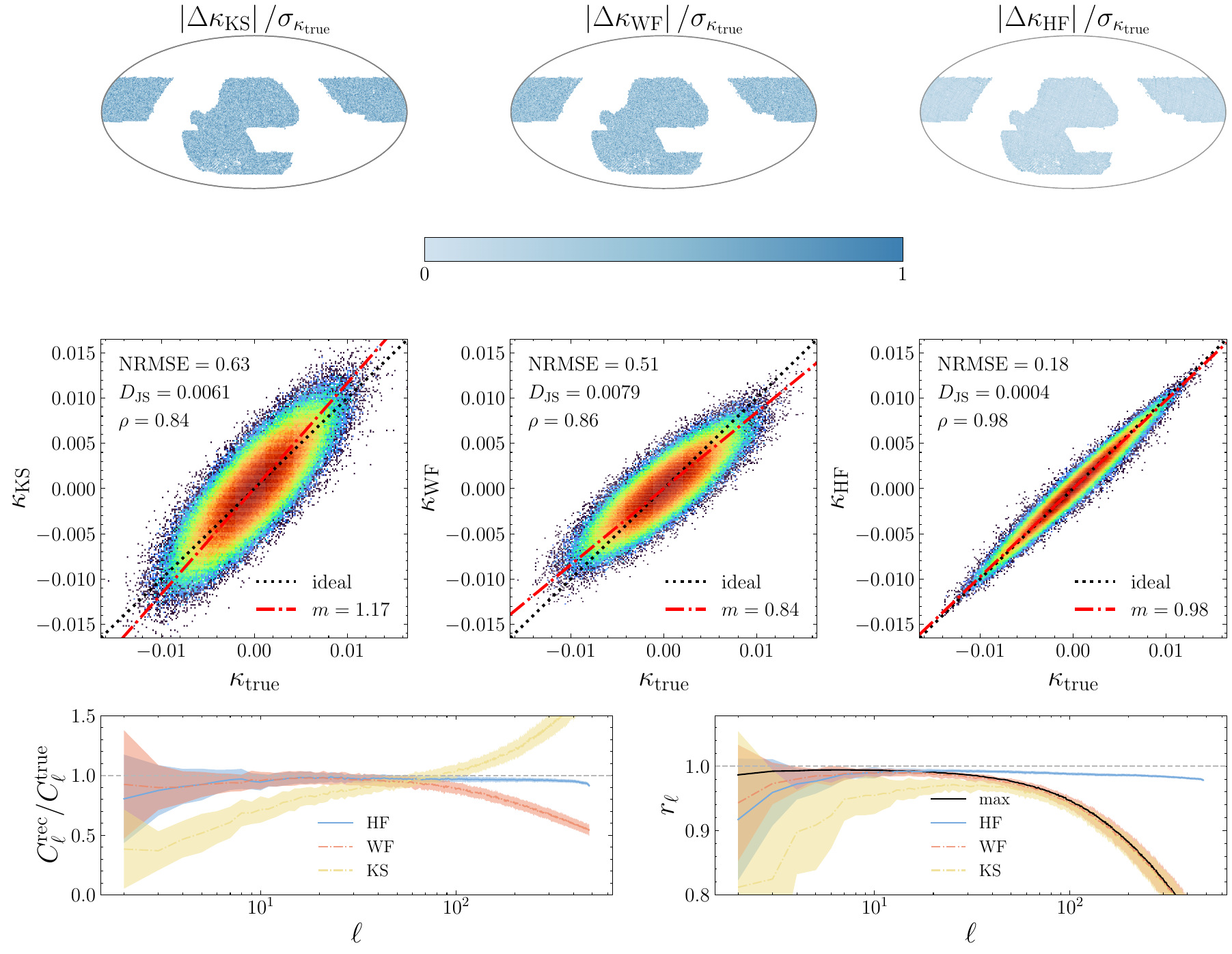}
  \caption{Comparison of reconstruction methods KS, WF, and HF on masked noisy maps at $N_{\mathrm{side}}=256$.\@ Top: normalized residuals $\Delta\kappa/\sigma_{\kappa_{\rm true}}$; lighter colors indicate smaller errors. Middle: pixelwise $\kappa_{\rm rec}$ versus $\kappa_{\rm true}$ with the diagonal denoting perfect recovery. WF reduces scatter relative to KS but is biased; HF attains tighter scatter with near-unity slope. Bottom: harmonic-space metrics, the power-spectrum ratio $C_\ell^{\mathrm{rec}}/C_\ell^{\mathrm{true}}$, and the cross-correlation $r_\ell$. The black curve shows the linear ceiling $r_\ell^{\max}$ from Eq.~(\ref{eq:r_max}).}\label{fig:noise_effect}
\end{figure*}

Noise imposes a fundamental limit distinct from masking. Missing data may be partially inferred from surrounding regions, whereas additive noise irreversibly degrades information, most severely on small angular scales. We first establish the linear theoretical limit, then assess how the learned model exceeds it in practice.

Consider isotropic, uncorrelated pixel noise in the observed shear $E$-mode coefficients,
\begin{equation}
  \hat{\gamma}^{E}_{\ell m}=\gamma^{E}_{\ell m}+n^{\gamma}_{\ell m},
  \qquad
  \langle|n^{\gamma}_{\ell m}|^2\rangle \equiv N^{\gamma}_\ell .
\end{equation}
On the full sky, the shear and convergence are related in harmonic space by
\begin{equation}\label{eq:shear_kappa_kernel}
  \gamma^{E}_{\ell m}=D_\ell\,\kappa_{\ell m},
  \qquad
  D_\ell \equiv -\sqrt{\frac{(\ell-1)(\ell+2)}{\ell(\ell+1)}} ,
\end{equation}
where the overall sign convention is irrelevant for power spectra and correlation coefficients.

It is convenient to work in ``$\kappa$-equivalent'' units by defining the rescaled mode
\begin{equation}\label{eq:kappa_equiv_shear}
  \hat{\gamma}_{\ell m}\equiv D_\ell^{-1}\hat{\gamma}^{E}_{\ell m}
  = \kappa_{\ell m}+n_{\ell m},
  \qquad
  n_{\ell m}\equiv D_\ell^{-1}n^{\gamma}_{\ell m}.
\end{equation}
In these units, the signal and noise spectra are defined by
\begin{equation}\label{eq:signal_noise_defs}
  C_{\ell}^{\mathrm{true}} \equiv \langle|\kappa_{\ell m}|^2\rangle,
  \qquad
  N_\ell \equiv \langle|n_{\ell m}|^2\rangle
  = \frac{N^{\gamma}_\ell}{|D_\ell|^2}.
\end{equation}
By construction, $\langle|\gamma_{\ell m}|^2\rangle=C_\ell^{\mathrm{true}}$ because $\gamma_{\ell m}$ in
Eq.~\eqref{eq:kappa_equiv_shear} denotes the rescaled field, not the physical shear mode $\gamma^{E}_{\ell m}$.

Any linear estimator can then be written with a multipole weight $\omega_\ell$ as
\begin{equation}\label{eq:estimator}
  \hat{\kappa}_{\ell m}
  = \omega_\ell \hat{\gamma}_{\ell m}
  = \omega_\ell\left(\kappa_{\ell m}+n_{\ell m}\right).
\end{equation}
The reconstructed auto-spectrum is
\begin{equation}
  C_\ell^{\mathrm{rec}}
  = \langle|\hat{\kappa}_{\ell m}|^2\rangle
  = \omega_\ell^2\big(C_\ell^{\mathrm{true}}+N_\ell\big),
\end{equation}
and the cross-spectrum with the truth is
\begin{equation}
  C_\ell^{\mathrm{rec,true}}
  = \langle \hat{\kappa}_{\ell m}\kappa^*_{\ell m}\rangle
  = \omega_\ell C_\ell^{\mathrm{true}},
\end{equation}
using independence of signal and noise. Inserting these into the definition of $r_\ell$ [Eq.~(\ref{eq:coefficient})] gives
\begin{align}\label{eq:r_max}
r_\ell
&= \frac{C_\ell^{\mathrm{rec,true}}}{\sqrt{C_\ell^{\mathrm{rec}}\,C_\ell^{\mathrm{true}}}}
= \frac{\omega_\ell C_\ell^{\mathrm{true}}}
       {\sqrt{\omega_\ell^2\bigl(C_\ell^{\mathrm{true}}+N_\ell\bigr)C_\ell^{\mathrm{true}}}}
\nonumber\\
&= \frac{\omega_\ell C_\ell^{\mathrm{true}}}
        {\omega_\ell\sqrt{\bigl(C_\ell^{\mathrm{true}}+N_\ell\bigr)C_\ell^{\mathrm{true}}}}
\qquad (\omega_\ell>0)
\nonumber\\
&= \sqrt{\frac{C_\ell^{\mathrm{true}}}{C_\ell^{\mathrm{true}}+N_\ell}}
\;\equiv\; r_\ell^{\max}.
\end{align}
Crucially, the filter weight $\omega_\ell$ cancels and any positive linear weight yields the same correlation coefficient. This value therefore represents the fundamental ceiling on phase recovery for \emph{all} linear estimators. When $N_\ell \gg C_\ell^{\mathrm{true}}$, the ceiling necessarily declines, independent of filter design.

Figure~\ref{fig:noise_effect} compares reconstruction methods on masked, noisy data at resolution $N_{\mathrm{side}}=256$ from the unseen test dataset. The real-space residual maps (top row) demonstrate systematic improvement across methods, with progressively lighter regions from KS to WF to HF indicating smaller and more spatially uniform errors. Quantitatively, the pixelwise regression analysis (middle row) confirms HF achieves near-unity slope ($m=0.98$) with minimal scatter, whereas WF reduces scatter relative to KS but introduces systematic bias ($m=0.84$) through regressive noise reduction. The harmonic-space diagnostics (bottom row) reveal complementary insights into reconstruction fidelity. KS systematically overestimates power in noise-dominated regimes, $C_\ell^{\mathrm{rec, KS}}/C_\ell^{\mathrm{true}}> 1$, while underestimating large-scale amplitudes. WF suppresses small-scale power through regularization and exhibits persistent bias for $\ell \gtrsim 100$. In contrast, HF maintains accurate amplitude recovery across all multipoles, with ratios closest to unity throughout the accessible $\ell$ range. Phase correlation coefficients exhibit similar behavior, where both KS and WF approach the linear theoretical ceiling $r_\ell^{\max}$, and KS shows additional degradation on large scales. At $N_{\mathrm{side}}=256$, this theoretical ceiling begins declining near $\ell \approx 40$ due to the noise term in Eq.~(\ref{eq:r_max}). To isolate noise-induced artifacts without confounding effects from pre/postprocessing, we present unsmoothed reconstructions in this comparison. Appendix~\ref{app:smoothing} examines smoothing as a commonly employed noise-reduction technique.

Linear estimators face a fundamental limit: smoothing can suppress noise in amplitude recovery ($C_\ell^{\mathrm{rec}}/C_\ell^{\mathrm{true}}$), but cannot restore phase information already corrupted by noise. HF circumvents this barrier by learning a nonlinear mapping that simultaneously denoises the field and recovers small-scale phases beyond the linear ceiling $r_\ell^{\max}$, while preserving nearly unbiased amplitudes across all multipoles. Our model is trained at a fixed noise level and therefore learns the statistical properties of that specific noise regime. For surveys with spatially varying or different noise characteristics, two strategies are available: (i) fine-tuning the pretrained model on simulations matching the target noise properties, or (ii) training on data spanning a range of galaxy densities to learn noise-invariant representations. The latter enables generalization across noise regimes without retraining.

\subsection{Border effects}

Reconstruction accuracy degrades near survey boundaries. Harmonic transforms and convolutions propagate values assigned to masked pixels into neighboring observed regions, creating systematic artifacts amplified by edge discontinuities \citep{changDarkEnergySurvey2018}. Existing solutions such as discrete cosine transforms \citep{piresEuclidReconstructionWeaklensing2020} and iterative optimization \citep{mawdsleyDarkEnergySurvey2020} provide limited improvement at substantial computational expense.

HF addresses this challenge through its asymmetric architecture. The encoder processes only visible and inner-edge patches, representing boundary pixels with learnable MASK tokens rather than imputed values. This design excludes distant masked regions from processing, limiting error propagation and enabling accurate reconstruction near survey edges.

\begin{figure}
  \centering
  \includegraphics[width=0.98\linewidth]{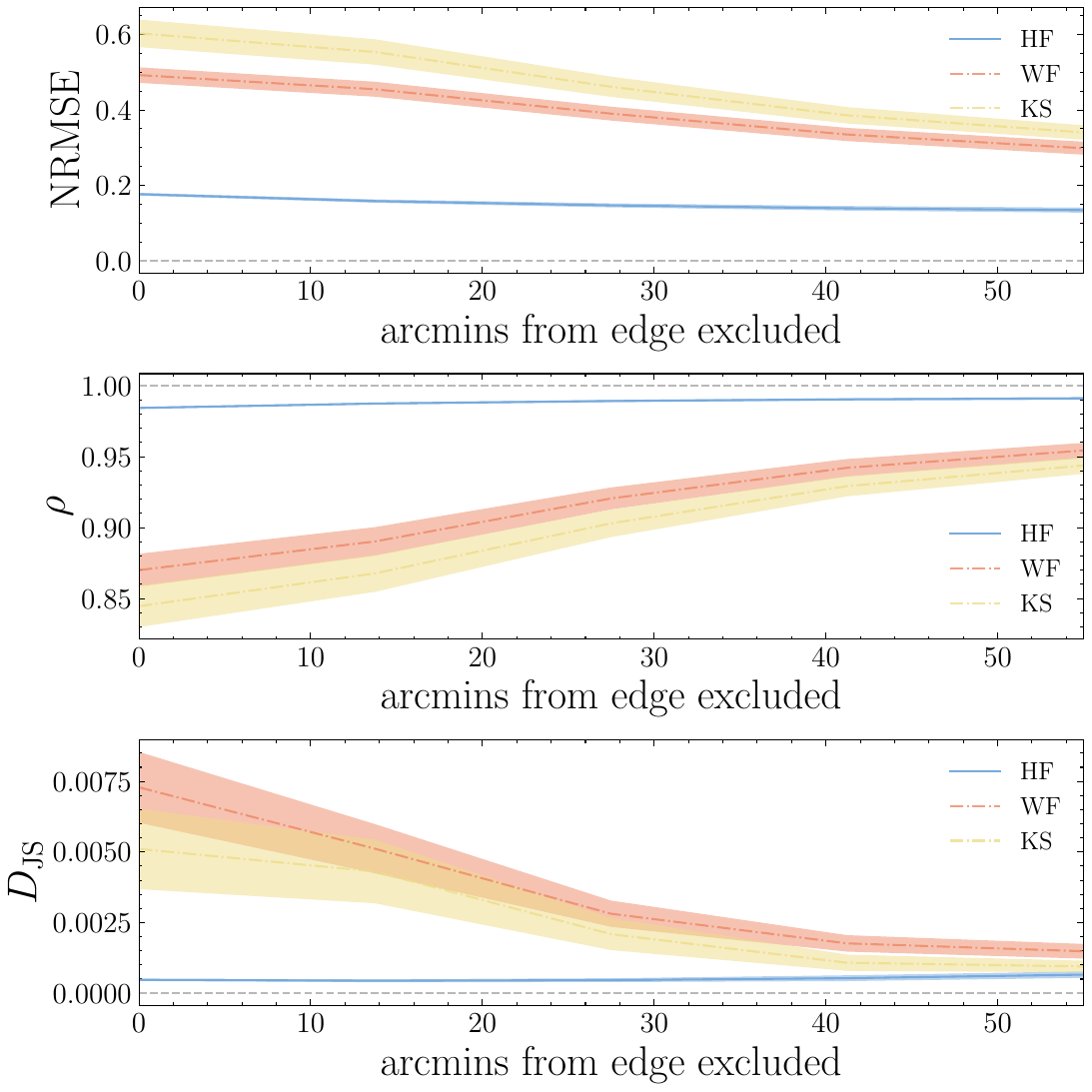}
  \caption{Quantitative performance as a function of edge exclusion width. We compute NRMSE, $\rho$, and $D_{\rm JS}$ on visible pixels after removing those within a specified distance of the mask boundary. HF remains accurate and stable across all exclusion widths, outperforming traditional methods that improve only when large edge regions are discarded.}\label{fig:border_effect_nside256}
\end{figure}

We quantify border effects following \citet{mawdsleyDarkEnergySurvey2020}. For each test map, we progressively exclude pixels within increasing distances of the mask boundary and evaluate NRMSE, $\rho$, and $D_{\rm JS}$ on the remaining visible pixels (Fig.~\ref{fig:border_effect_nside256}; also see Sec.~\ref{sec:data_generation}). KS and WF show marked gains as the exclusion width grows, revealing degraded accuracy near edges. HF attains higher accuracy with no exclusion and maintains nearly constant performance across all widths, which indicates effective suppression of border artifacts and reliable reconstruction close to survey edges.

\subsection{Rotation invariance}\label{sec:rotation}

Deep learning models on the sphere must remain accurate under arbitrary rotations, a harder requirement than translation invariance on flat maps. We enforce rotation invariance with on-the-fly random rotations of the masks during training and learnable positional embeddings that adapt to these transformations. We have tested that fixed (nonlearnable) embeddings fail to preserve accuracy under rotation.

As detailed in Appendix~\ref{app:rotation}, we test rotation invariance by applying random rotations to the DECaLS mask. Reconstruction quality remains stable across all angles, and pixelwise comparisons between the original and rotated cases are nearly identical. This demonstrates robust rotation invariance. In addition to rotations, the same HEALFormer model handles different survey footprints without retraining, as shown in Appendix~\ref{app:variable_sized_input}. 

\subsection{Generalization to different cosmologies}\label{sec:multiverse}

A central requirement for deep learning on weak lensing mass mapping is the ability to train once and perform reliably on data generated under different cosmological parameters. As described in Sec.~\ref{sec:data_generation}, earlier experiments trained HF on simulations drawn from the KiDS posterior and showed strong performance across diverse cosmologies (Sec.~\ref{sec:noise_effect}). To exclude the possibility that HF is merely interpolating within the training set, we train HF on a single cosmology fixed to the Planck~2018 parameters and evaluate it on simulations that span a broad range of models. The test set varies the matter density $\Omega_m$ and the fluctuation amplitude $\sigma_8$, with cosmologies sampled from the KiDS\mbox{-}450 posterior \citep{Hildebrandt2017}. If performance remains stable, this indicates that HF has learned the mapping from shear to convergence rather than exploiting interpolation in parameter space.

\begin{figure}
  \centering
  \includegraphics[width=0.98\linewidth]{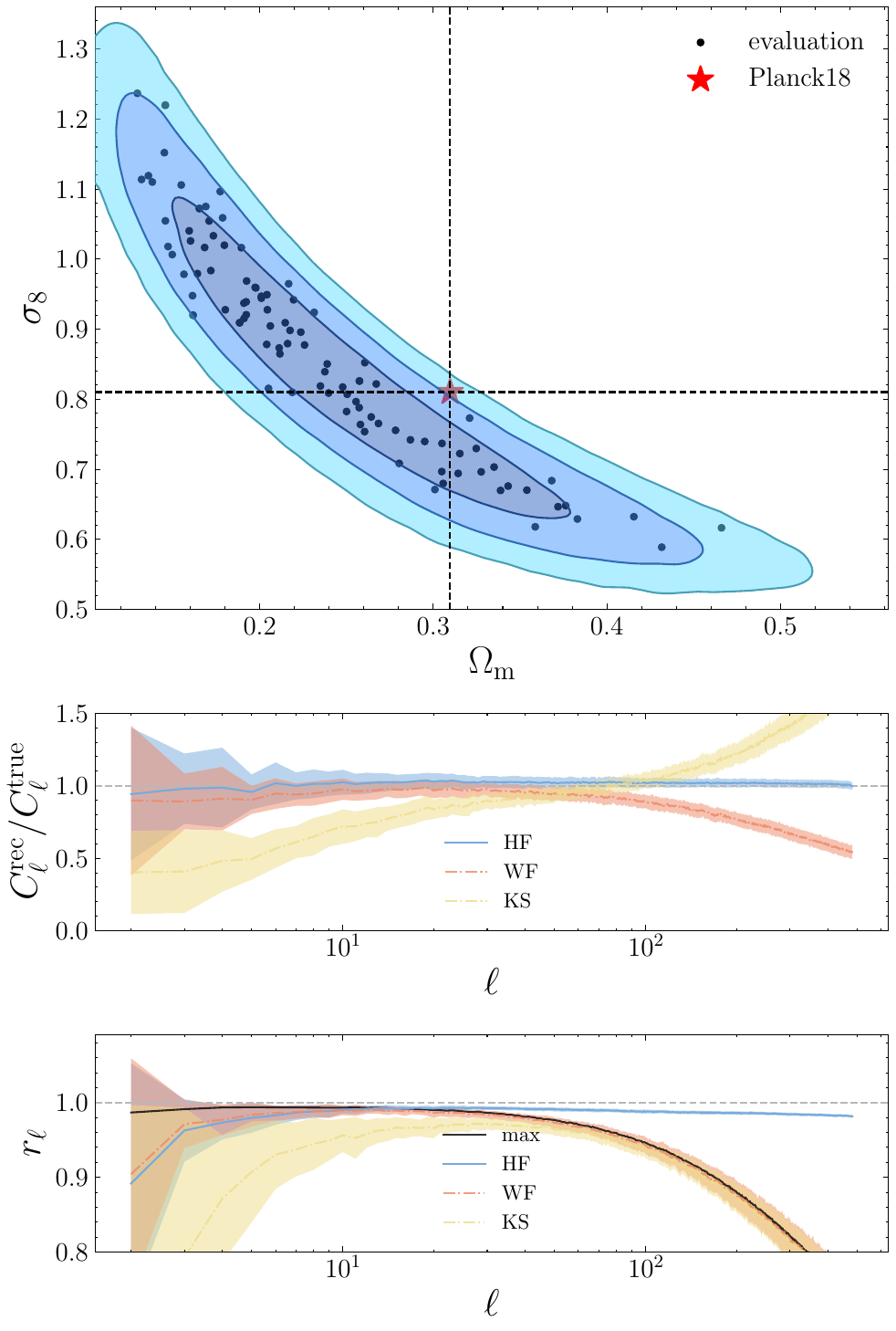}
  \caption{Generalization across cosmologies. Top: distribution of $(\Omega_m,\sigma_8)$ showing the training cosmology from Planck~2018 (red star), sampled test cosmologies (black points), and the KiDS posterior (blue contours). Middle: angular power-spectrum ratio $C_\ell^{\mathrm{rec}}/C_\ell^{\mathrm{true}}$ for HF (blue) versus the ideal target (black dotted). Bottom: cross-correlation coefficient $r_\ell$ between reconstructed and true fields. HF exceeds the linear\mbox{-}method ceiling $r_\ell^{\max}$ at small scales and matches WF performance at large scales $\ell \lesssim 10$.}\label{fig:multiverse}
\end{figure}

Figure~\ref{fig:multiverse} shows that a model trained only on Planck~2018 transfers cleanly to unseen cosmologies. The reconstructed amplitudes remain accurate across all test cosmologies, with $C_\ell^{\mathrm{rec}}/C_\ell^{\mathrm{true}} \approx 1$ over a wide range of scales. Phase recovery is similarly robust: $r_\ell$ surpasses the theoretical limit for linear estimators on small scales and matches WF performance at large scales. These results confirm the findings of Section~\ref{sec:noise_effect} and demonstrate out\mbox{-}of\mbox{-}distribution generalization beyond the training cosmology.
However, the learned prior remains tied to the training simulations. Whether the model maintains accuracy under alternative cosmologies, baryonic feedback, or extensions beyond $\Lambda$CDM is an open question for future investigation.

\subsection{Reconstruction at high resolution}\label{sec:denoising}

A central difficulty in high-resolution mass mapping is the larger per pixel noise that obscures small-scale structure. Linear estimators such as the WF reduce noise by aggressively surpassing amplitudes and do not recover phase information where noise dominates. It also requires accurate priors for signal and noise covariances and relies on iterative solves that are costly, for example $\sim 2$ hours for a single $N_{\mathrm{side}}=1024$ map. HF learns a nonlinear mapping from shear to convergence and avoids explicit covariance priors. Once trained, it performs inference in seconds, which enables pipelines that require many reconstructions such as simulation based inference.

\begin{figure*}
  \centering
  \includegraphics[width=0.9\textwidth]{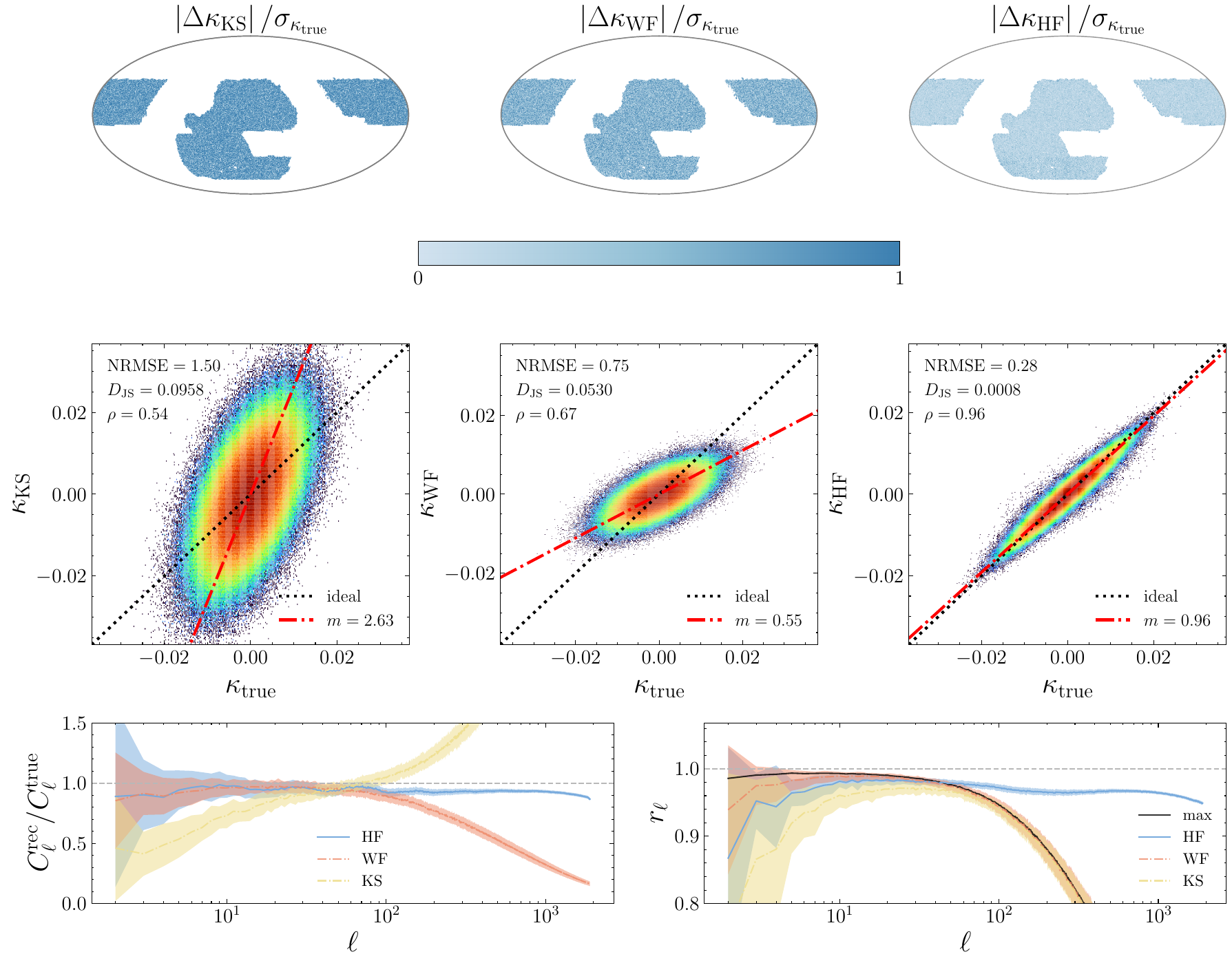}
  \caption{High-resolution comparison on noisy $N_{\mathrm{side}}=1024$ maps with the DECaLS mask. Top: residual maps of reconstruction error; KS shows the largest deviations, WF improves accuracy, and HF is the cleanest. Middle: pixelwise comparisons of reconstructed versus true convergence; HF has the smallest scatter and slope $m$ closest to unity, where $m$ is the fitted slope. Bottom: harmonic domain results showing power-spectrum ratios (left) and phase correlations (right). WF suppresses small-scale noise; HF maintains the best performance across all scales, matches WF on large scales ($\ell \lesssim 10$), and exceeds the phase recovery limit at small scales (black line).}\label{fig:denoising_nside1024}
\end{figure*}

Figure~\ref{fig:denoising_nside1024} presents high-resolution results at $N_{\mathrm{side}}=1024$, where per-pixel noise exceeds that of the $N_{\mathrm{side}}=256$ case (Fig.~\ref{fig:noise_effect}) and increases scatter in the pixelwise comparison. Despite this challenge, HF trained with the coarse-to-fine strategy (Sec.~\ref{sec:coarse-to-fine}) produces the cleanest residual maps and smallest scatter among all methods, demonstrating effective denoising. The reconstruction is nearly unbiased in pixel space ($m \approx 1$), and the one-point distribution of the reconstructed convergence tracks the truth more closely than KS or WF (Appendix~\ref{app:PDF}). In harmonic space, HF matches WF at large scales while exceeding the theoretical phase-recovery bound $r_\ell^{\max}$ at small scales.

These results reproduce the trends at lower resolution in Fig.~\ref{fig:noise_effect}. HF scales robustly to higher resolution, addresses the denoising challenge, and meets throughput constraints, with training feasible on a single GPU.\@ 

\subsection{Painting at high resolution}

\begin{figure}
  \centering
  \includegraphics[width=0.98\linewidth]{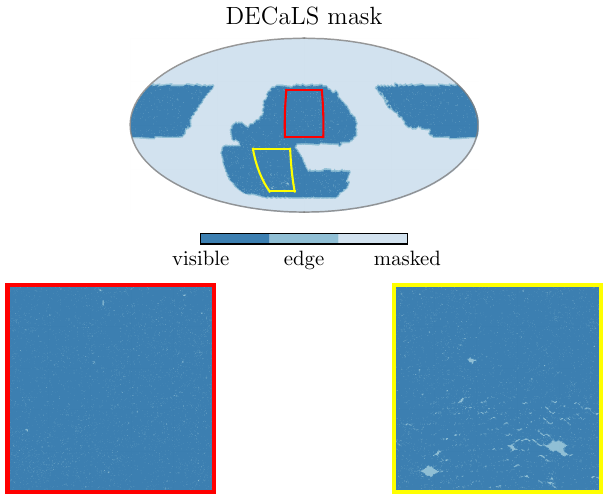}
  \caption{DECaLS mask at $N_{\mathrm{side}}=1024$ with two $40\degree \times 40\degree$ regions marked. The full-sky view separates observed pixels (dark blue), masked pixels along the survey boundary (light blue), and distant masked pixels excluded from analysis (grey). The red square contains masked points, and the yellow square contains extended masked holes.}\label{fig:zoom_mask_nside1024}
\end{figure}

\begin{figure*}
  \centering
  \includegraphics[width=0.8\textwidth]{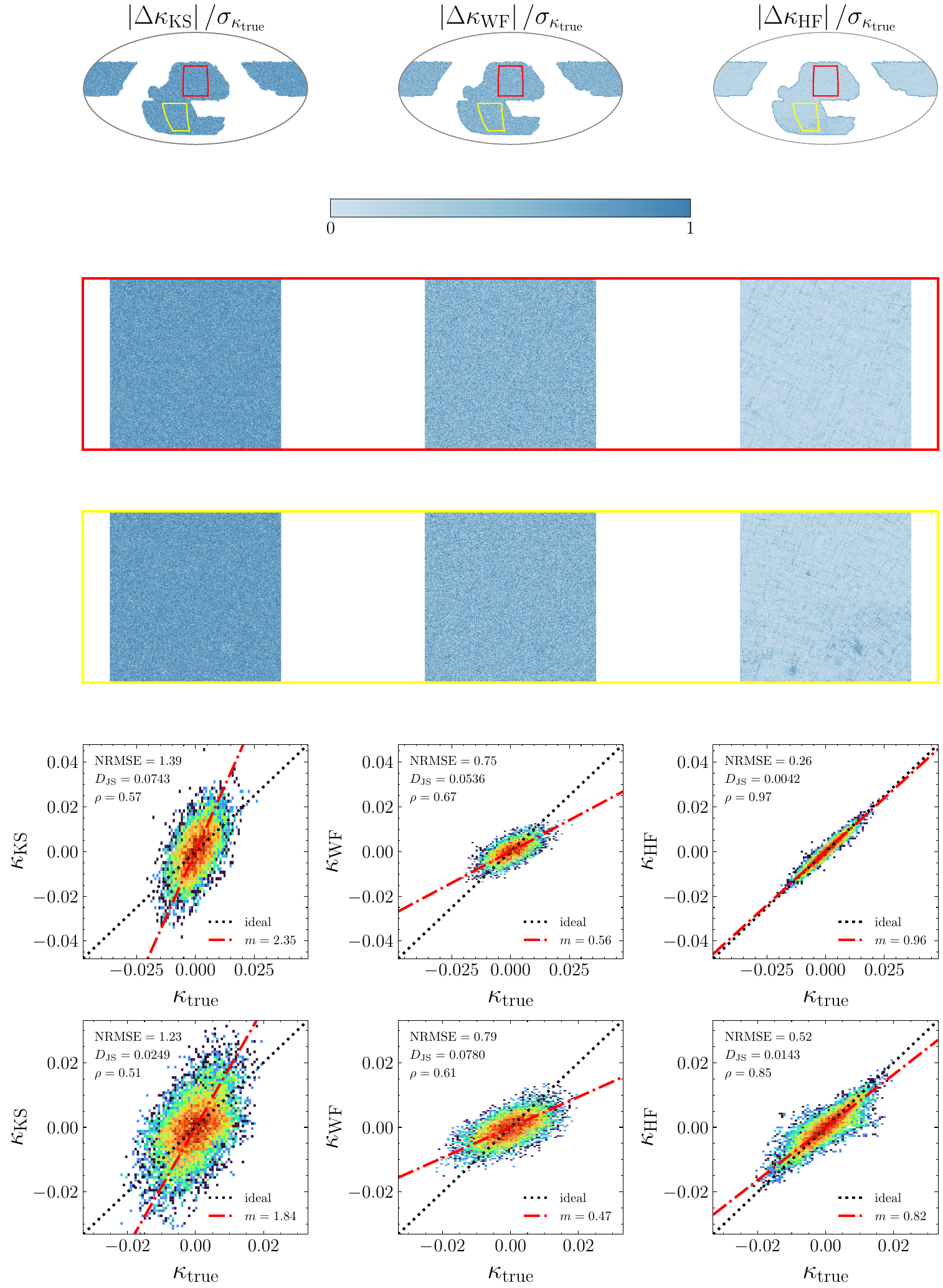}
  \caption{Reconstruction quality across masked pixels within the two selected DECaLS regions at $N_{\mathrm{side}}=1024$. The top row displays residual maps for each method. The middle rows present magnified views of the pointlike (red box) and large-hole (yellow box) regions marked in Fig.~\ref{fig:zoom_mask_nside1024}, showing both visible and inner-edge pixels that were unseen during reconstruction. The bottom rows provide pixelwise comparisons between reconstructed and true convergence for inner-edge pixels only. In the point-like region, HF achieves nearly unbiased reconstruction with slope $m \approx 1$ and minimal scatter. Although the large-hole region presents greater difficulty due to its extended masked area, HF maintains superior performance with the smallest residuals and tightest correlation relative to competing methods. Faint grid artifacts visible in the HF residual maps arise from imperfect denoising but remain subdominant, confirming that HF delivers the highest reconstruction fidelity across both masked geometries.}\label{fig:painting_nside1024}
\end{figure*}

Painting reconstructs values in edge pixels near survey boundaries by using nonlocal information from surrounding observations. In HF this is implemented with a learnable MASK token that encodes masked pixels or patches. Unlike zero filling in earlier work such as DeepMass, this representation avoids confusing missing data with true zeros. Training uses an adaptive weighted loss to balance contributions from visible, inner-edge, and distant masked pixels (Sec.~\ref{sec:loss_function}).

We evaluate painting within the DECaLS footprint at $N_{\mathrm{side}}=1024$ using the two $40\degree \times 40\degree$ regions highlighted in Fig.~\ref{fig:zoom_mask_nside1024}: one dominated by pointlike (red) and one with extended masked holes (yellow). Note that the mass mapping is performed on the sphere and these square cutouts are shown to visualize the reconstructions.

Figure~\ref{fig:painting_nside1024} summarizes the results. Residual maps reveal that HF produces the cleanest reconstructions, with faint grid patterns reflecting imperfect denoising that remain subdominant. Pixelwise tests of masked pixels in the pointlike region exhibit near-unity slope ($m \approx 1$) with minimal scatter, maintaining consistency with the performance achieved in visible pixels (Sec.~\ref{sec:noise_effect}). The large-hole region presents greater difficulty, where scatter increases and the slope degrades to $m=0.82$. Nevertheless, HF maintains superior performance relative to all competing methods across both masked geometries. The reconstruction quality achieved in these diverse configurations confirms that the model performs robust painting across realistic survey footprints with varying masked-region morphologies.

\section{Conclusion}\label{sec:conclusion}

We have presented HEALFormer, a transformer-based architecture for weak lensing mass mapping that operates natively on the HEALPix sphere and accommodates realistic survey footprints from DES, KiDS, DECaLS, and Planck through learnable mask tokens. Trained with shape noise at LSST-Y10/Euclid levels ($n_{\rm gal}=30\,\mathrm{arcmin}^{-2}$), HEALFormer outperforms Kaiser-Squires and Wiener filter reconstructions at both $N_{\mathrm{side}}=256$ and $1024$, achieving lower normalized RMSE, higher Pearson correlation $\rho$, and reduced Jensen-Shannon divergence $D_{\rm JS}$. Residual maps and one-to-one comparisons reveal minimal bias with reduced scatter relative to classical methods. In harmonic space, the cross-correlation coefficient $r_\ell$ exceeds the linear-method ceiling $r_\ell^{\max}$, demonstrating phase recovery beyond what linear approaches can achieve. Unlike the Wiener filter, HEALFormer learns the shear-to-convergence mapping directly from data without requiring explicit covariance matrices, though the current implementation assumes a fixed noise level that is implicitly encoded in the learned weights. Parameter-efficient fine-tuning via LoRA enables training and inference on a single GPU while maintaining high reconstruction fidelity across diverse survey geometries and mitigating border effects. The model generalizes to cosmologies not encountered during training, confirming that it captures the underlying physical transformation rather than memorizing training examples. By recovering both amplitude and phase with unprecedented accuracy and exceeding linear-theory limits at small scales, HEALFormer establishes a practical, efficient, and physically principled approach to spherical mass mapping for current and next-generation wide-field surveys.

The architecture embodies three key advances that address fundamental challenges in spherical weak lensing analysis. It operates natively on the sphere by processing HEALPix inputs with learnable position embeddings that encode sky geometry and avoid flat-sky approximations. It scales efficiently through a coarse-to-fine training strategy combined with LoRA fine-tuning, enabling high-resolution reconstruction at $N_{\mathrm{side}}=1024$ on standard hardware. It handles complex survey geometries through learnable mask tokens that explicitly represent missing data, allowing principled inpainting and robust recovery across irregular footprints.

Several directions remain open.
The present work assumes a fixed $\Lambda$CDM cosmology and idealized simulation conditions; extending to alternative cosmological models (e.g., dynamical dark energy, modified gravity) and incorporating baryonic physics (e.g., AGN feedback) are important avenues. The current model also learns an implicit noise-level prior from the training data; extending to non-Gaussian and spatially correlated noise components would better capture realistic observational conditions. Reducing patch size may improve reconstruction fidelity at increased computational cost. Developing controlled inpainting strategies beyond survey footprints, analogous to constrained realizations in the Wiener filter, could expand the effective reconstruction area. We will explore these extensions in future work.

This research leveraged several essential open-source scientific computing tools. For core numerical computations, we utilized NumPy \citep{harrisArrayProgrammingNumPy2020} and SciPy \citep{virtanenSciPy10Fundamental2020}. The spherical data manipulation was performed using \texttt{HEALPix/healpy} \citep{GORSKI2005, zoncaHealpyEqualArea2019}, while cosmological calculations relied on CCL \citep{chisariCoreCosmologyLibrary2019}. Our deep learning implementation was built on PyTorch \citep{anselPyTorch2Faster2024} and the Transformers \citep{wolfHuggingFacesTransformersStateoftheart2020}.

\section*{Acknowledgements}
This work was supported by the National Key R\&D Program of China (No. 2023YFA1607800, No. 2023YFA1607801, No. 2023YFA1607802), the National Science Foundation of China (Grant No. 12595310, No. 12273020), the China Manned Space Project with No. CMS-CSST-2021-A03 and No. CMS-CSST-2025-A04, the ``111'' Project of the Ministry of Education under Grant No. B20019, and the sponsorship from Yangyang Development Fund.
This project is supported in part by Office of Science and Technology, Shanghai Municipal Government (Grant No. 24DX1400100, No. ZJ2023-ZD-001).
This work made use of the Gravity Supercomputer at the Department of Astronomy, Shanghai Jiao Tong University, which 
enabled the large-scale experiments presented in this work.

\section*{DATA AVAILABILITY}

The data that support the findings of this article are openly available \cite{healformergithub},
embargo periods may apply.


\clearpage
\newpage
\appendix

\section{IMPLEMENTATION DETAILS AND
ADDITIONAL RESULTS}

\subsection{Position embedding}\label{app:position_embedding}

Self-attention [Eq.~(\ref{eq:attention})] is permutation invariant: swapping two input tokens leaves the output unchanged. To encode the geometry of the sphere, we add explicit positional information to the patch tokens so the model can learn spatial relationships on the HEALPix grid.

We consider (i) fixed sinusoidal embeddings and (ii) learnable embeddings. The fixed scheme of \citet{vaswaniAttentionAllYou2023} assigns a deterministic vector to each position index $pos$:
\begin{align}
  \mathrm{PE}_{pos,\,2i}   &= \sin\!\left(\frac{pos}{10000^{\,2i/d_{\rm model}}}\right),\\
  \mathrm{PE}_{pos,\,2i+1} &= \cos\!\left(\frac{pos}{10000^{\,2i/d_{\rm model}}}\right),
\end{align}
with $i=0,\dots,(d_{\rm model}/2-1)$. This choice is parameter-free and exposes relative offsets through inner products. Learnable embeddings instead assign a trainable vector to each position \citep{gehringConvolutionalSequenceSequence2017}, as in BERT \citep{devlinBERTPretrainingDeep2019}. In both cases we add the positional vector to the patch embedding before the Transformer blocks.

HEALPix with NEST ordering provides a natural 1D sequence; following \citet{dosovitskiyImageWorth16x162020}, we therefore adopt 1D positional indices rather than constructing an explicit 2D scheme. This retains HEALPix's hierarchical neighborhood structure while keeping the embedding simple.

\begin{figure}
  \centering
  \includegraphics[width=0.98\linewidth]{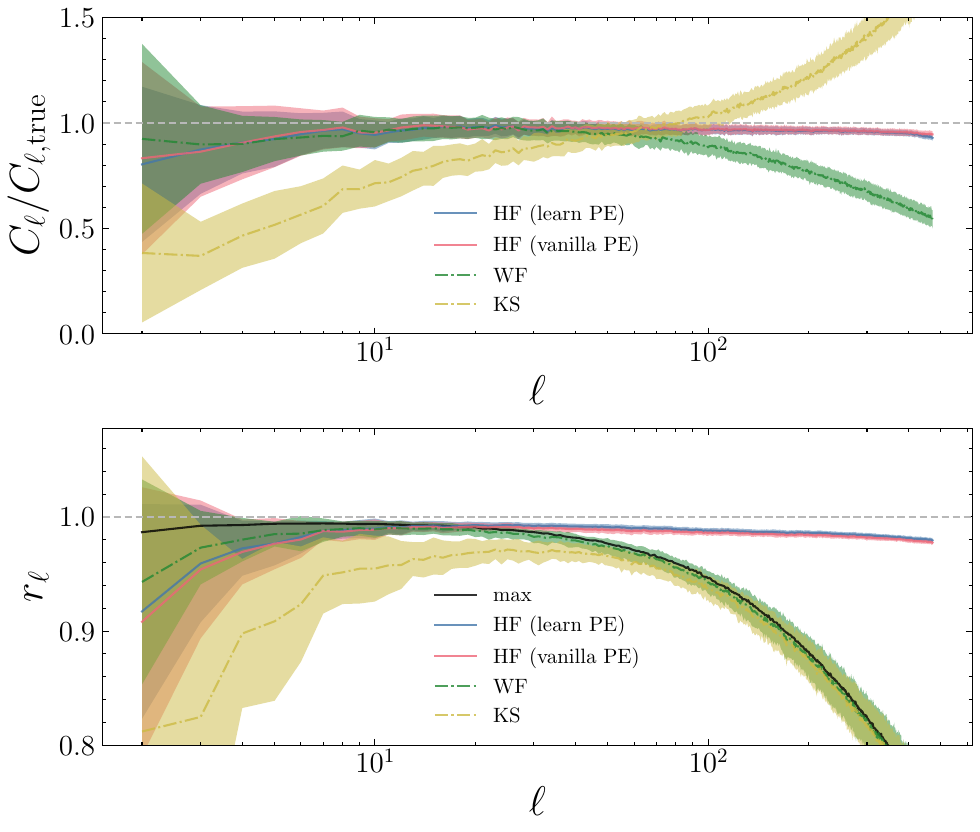}
  \caption{Fixed (sinusoidal) vs.\ learnable position embeddings on the sphere at $N_{\mathrm{side}}=256$ with the DECaLS mask. Top: ratio of reconstructed to true $C_\ell$; bottom: cross-correlation coefficient $r_\ell$. For fixed masks the two choices are statistically indistinguishable, but only learnable embeddings remain stable under random rotations.}\label{fig:vanillaPE_vs_learnPE}
\end{figure}

On fixed survey masks, fixed and learnable embeddings perform essentially identically in both amplitude and phase recovery (Fig.~\ref{fig:vanillaPE_vs_learnPE}, top and bottom panels). The difference appears when we randomly rotate the mask-map pairs during training and evaluation. Fixed sinusoidal embeddings fail to converge reliably under these rotations, whereas learnable embeddings adapt and maintain performance. In practice, if rotation invariance is not required, fixed embeddings are an adequate, lightweight choice; when rotation augmentation is used, learnable embeddings are necessary.

\begin{figure}
  \centering
  \includegraphics[width=0.98\linewidth]{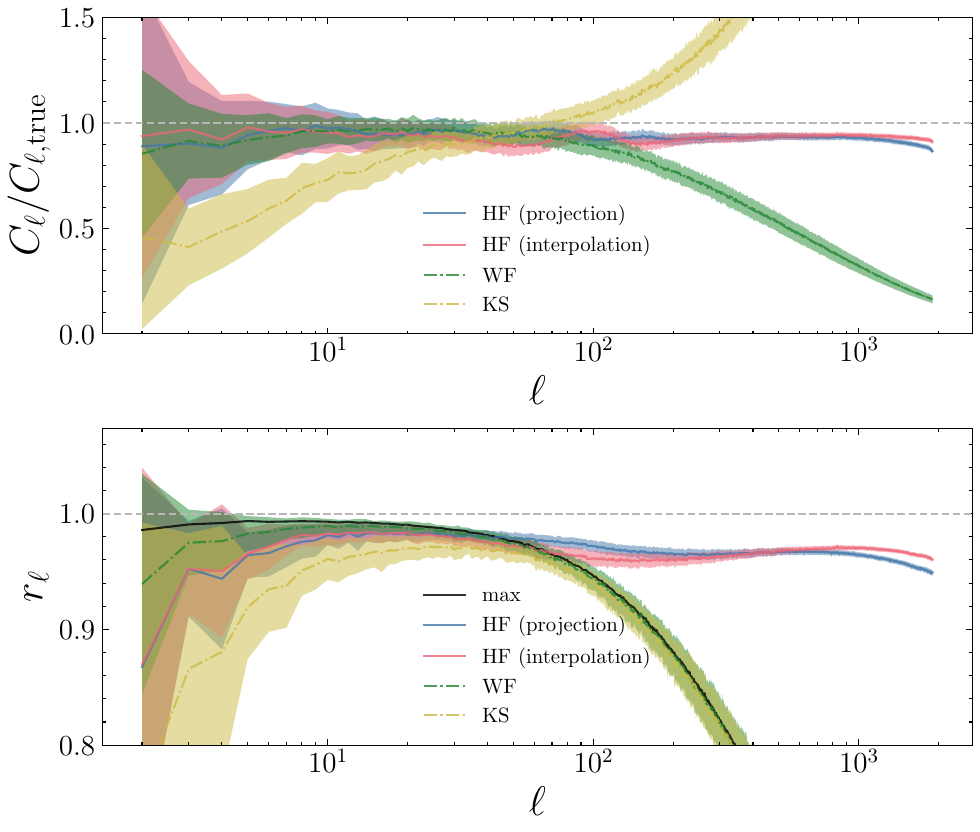}
  \caption{Upscaling learnable position embeddings from coarse to fine grids at $N_{\mathrm{side}}=1024$. A learned projection from coarse to fine positions (blue) is stable and preserves both amplitude (top) and phase (bottom). Simple bilinear interpolation (red) induces a wiggle around $\ell\!\sim\!100$ and lowers $r_\ell$.}\label{fig:projection_vs_interpolation}
\end{figure}

To reach $N_{\mathrm{side}}=1024$ efficiently we train coarsely and then fine-tune at high resolution. This requires lifting low-resolution positional vectors to the finer grid. We tested two strategies:
(i) \emph{projection}, a learned mapping from coarse to fine positions; and
(ii) \emph{bilinear interpolation}, a parameter-free upsampling.
Interpolation introduces a pronounced wiggle in the power-spectrum ratio near $\ell\!\sim\!100$ and degrades $r_\ell$, whereas projection is stable across scales (Fig.~\ref{fig:projection_vs_interpolation}). We therefore adopt projection when increasing resolution.

Additive position embeddings are essential for spherical mass mapping with transformers. Fixed sinusoidal embeddings suffice for static mask geometry; learnable embeddings are preferred when rotation augmentation or geometry adaptation is required. For high-resolution applications, projecting coarse learnable embeddings to the fine grid avoids spectral artifacts and preserves reconstruction fidelity.

\subsection{Reconstruction performance in the noise-free limit}\label{app:mask_effect}

\begin{figure*}
  \centering
  \includegraphics[width=0.98\textwidth]{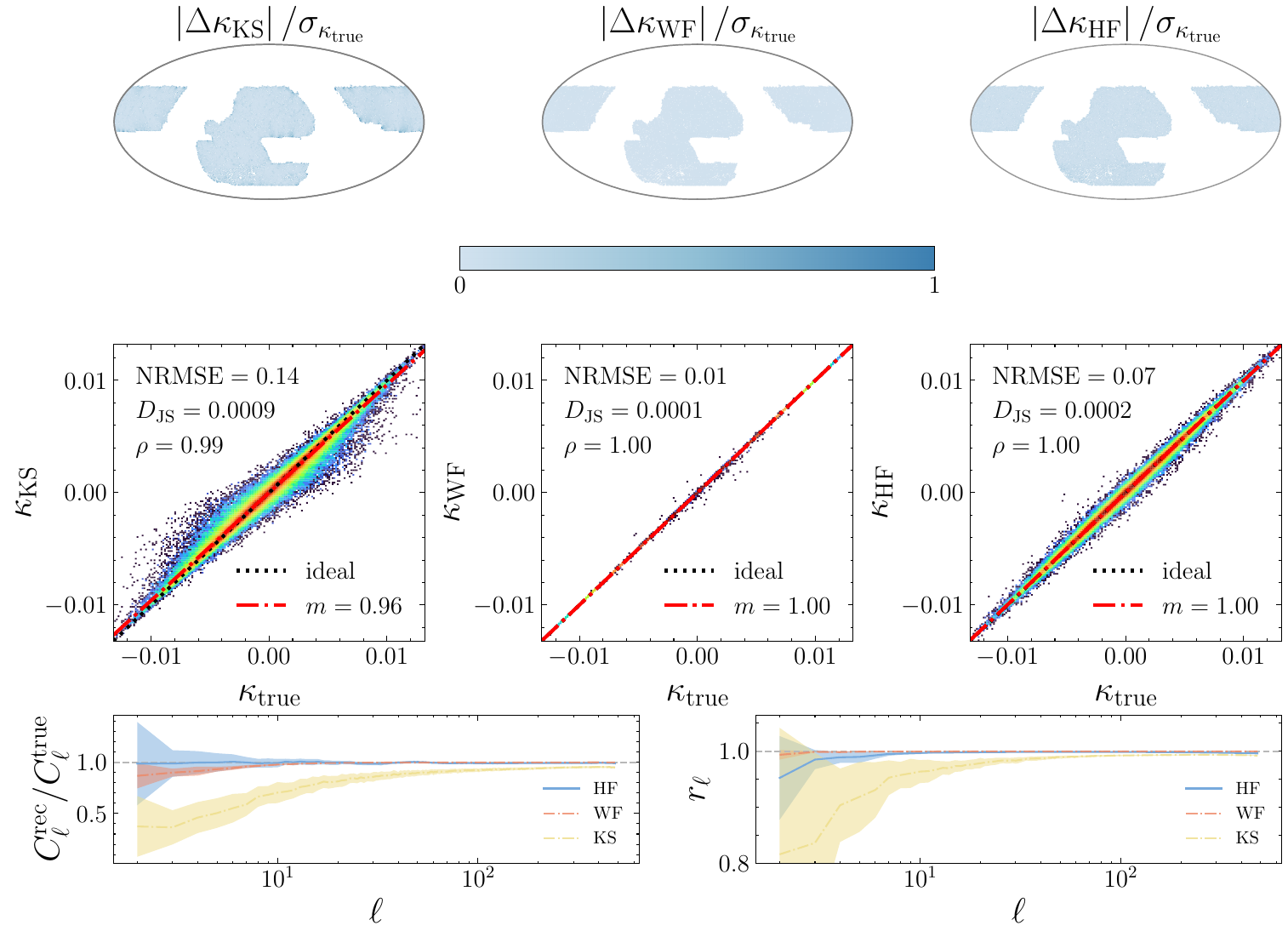}
  \caption{Comparison of reconstruction methods applied to masked clean maps at $N_{\mathrm{side}}=256$ without shape noise. Top row: normalized residual maps showing spatial distribution of reconstruction errors. Middle row: Pixel-by-pixel comparison between reconstructed and true convergence values, with perfect reconstruction represented by the diagonal line. Bottom row: harmonic-space evaluation showing angular power-spectrum ratio and cross-correlation coefficient ($r_{\ell}$), with the dashed grey line indicating theoretical perfect phase reconstruction. WF demonstrates near-perfect reconstruction in both pixel and harmonic space, KS shows significant degradation at large scales, while HF achieves high fidelity across all scales.}\label{fig:mask_effect} 
\end{figure*}

Here, we assess the impact of survey masks on convergence reconstruction in the absence of shape noise. This idealized scenario isolates the effect of missing data from observational noise. In most areas, when typical survey masks remain largely contiguous, the information contained within the observed region suffices for accurate recovery of the convergence field.

The three methods employ fundamentally different strategies to handle masked pixels. KS fills masked regions with zeros, which introduces biased reconstruction. WF assigns infinite noise variance to masked pixels, effectively excluding them from the reconstruction while solving for their values through the spatial covariance structure. HF represents masked pixels explicitly through learnable MASK tokens that enable the network to inpaint missing regions based on patterns learned during training.

From an information-theoretic perspective, near-perfect convergence recovery should be achievable in the noise-free masked case. For WF, the solution in Eq.~(\ref{eq:wf_solution}) remains well-defined even when the matrix $(\mathbf{S}_{\kappa}^{-1}+\mathbf{A}^\dagger\mathbf{N}^{-1}\mathbf{A})$ lacks full rank, since pseudoinverse methods apply. With contiguous survey footprints, the effective system approaches full rank, permitting near-perfect phase reconstruction ($r_\ell=1$) as demonstrated by AKRA~\citep{shiAccurateKappaReconstruction2024,shiAKRA20Accurate2024}.

Figure~\ref{fig:mask_effect} confirms these expectations. WF achieves near-perfect reconstruction in both pixel and harmonic space because assigning infinite noise variance to masked pixels eliminates dependence on the signal covariance, leaving only the precisely known noise covariance to control the solution. KS exhibits substantial degradation at large scales, unable to distinguish genuinely masked regions from true zero-valued signal. HF approaches the WF result but does not fully match it. This residual discrepancy likely reflects the fact that noise-free masked reconstruction is fundamentally linear, whereas the HF architecture has been optimized to capture nonlinear mappings that dominate in realistic noise-contaminated scenarios.

\subsection{Smoothing to reduce noise}\label{app:smoothing}
\begin{figure}
  \centering
  \includegraphics[width=0.98\linewidth]{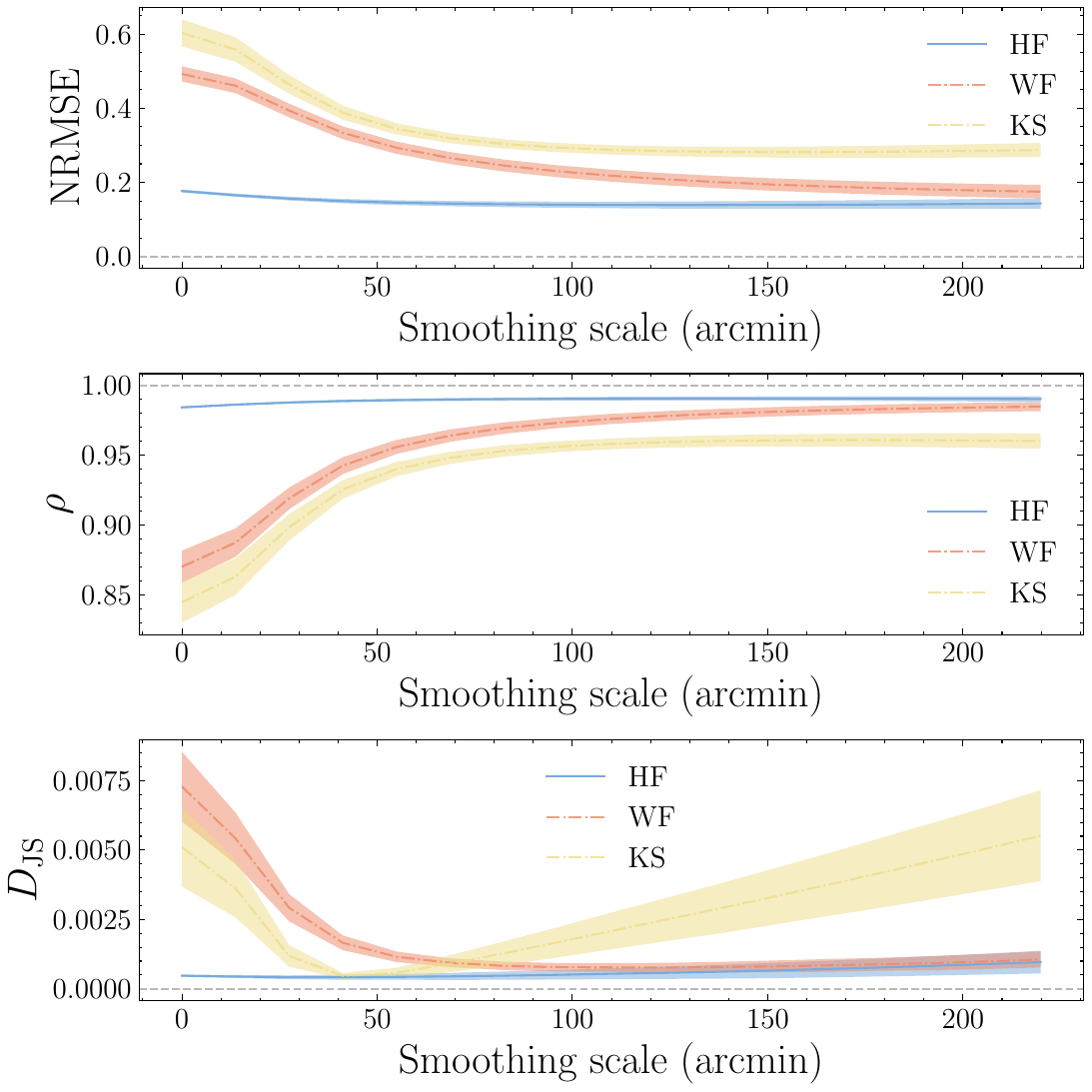}
  \caption{Performance comparison of reconstruction methods across different smoothing scales in the visible region. The dashed black line represents the perfect reconstruction limit. HF (blue line) demonstrates superior performance across all evaluation metrics and the entire range of smoothing scales.}\label{fig:smoothing_nside256}
\end{figure}

A common approach to mitigate noise in weak lensing maps is to apply smoothing. We evaluate the reconstruction performance of different methods under varying smoothing scales, as illustrated in Fig.~\ref{fig:smoothing_nside256}. The smoothing is implemented using a Gaussian kernel, characterized by its full width at half maximum. For this analysis, we employ a simplified approach without mask apodization. To prevent power leakage effects in the smoothing process, masked pixels are set to zero, and our comparison focuses exclusively on the visible region. Across all quantitative metrics (NRMSE, $\rho$, $D_{\rm JS}$), HEALFormer consistently outperforms other methods. This superior performance persists even at large smoothing scales ($\sim 200$ arcmin) where noise is effectively eliminated. Furthermore, HEALFormer exhibits remarkable stability in its performance across the entire range of smoothing scales, demonstrating its robustness to varying levels of map smoothing.

\subsection{PDF of reconstructed convergence fields}\label{app:PDF}
\begin{figure*}
  \centering
  \includegraphics[width=0.8\textwidth]{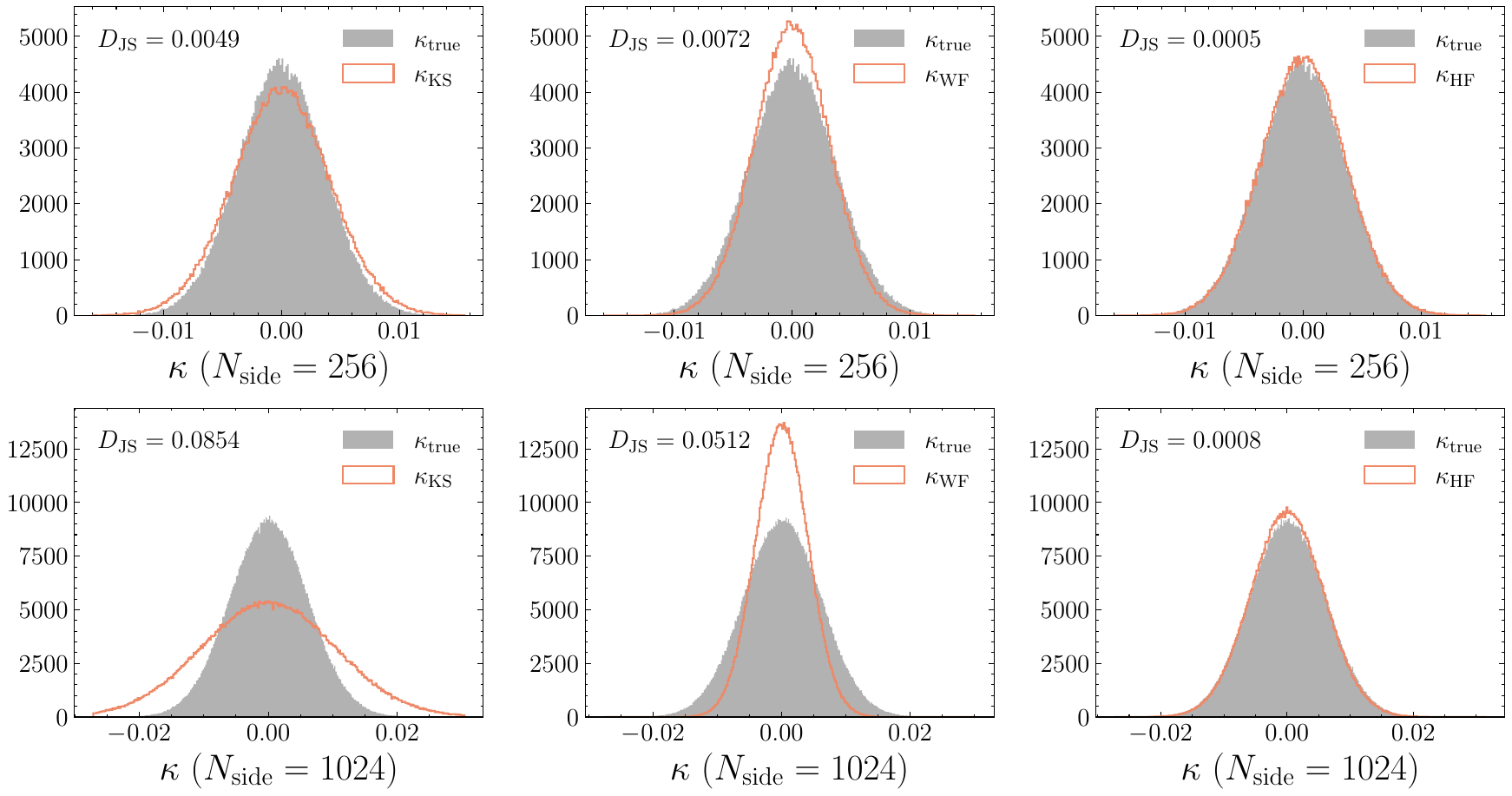}
  \caption{One-dimensional PDFs of reconstructed convergence fields in visible regions at low ($N_{\mathrm{side}}=256$, top row) and high resolution ($N_{\mathrm{side}}=1024$, bottom row). The panels compare KS (left), WF (middle), and HF (right) reconstructions against the true convergence distribution (black). The Jensen-Shannon divergence ($D_{\rm JS}$) quantifies the similarity between reconstructed and true distributions, with $D_{\rm JS}=0$ indicating perfect agreement. At high resolution, increased noise significantly degrades KS and WF performance compared to low resolution results. While WF's aggressive noise suppression leads to an overly compressed distribution, HF maintains excellent fidelity to the true distribution at both resolutions, as evidenced by its consistently low $D_{\rm JS}$ values.}\label{fig:PDF}
\end{figure*}

To further evaluate reconstruction quality, we analyze the statistical properties of the recovered convergence fields through their one-dimensional probability distribution functions (PDFs). Figure~\ref{fig:PDF} presents a comprehensive comparison using the Jensen-Shannon divergence ($D_{\rm JS}$) as a quantitative measure of distribution similarity. The analysis reveals that increased noise at high resolution severely impacts both KS and WF reconstructions, with WF's inherent noise suppression resulting in an artificially narrow distribution. In contrast, HEALFormer demonstrates remarkable consistency across both resolution regimes, preserving the true convergence distribution with high fidelity. This superior performance in maintaining accurate statistical properties further validates HEALFormer's effectiveness as a robust reconstruction method.

\subsection{Rotation invariance}\label{app:rotation}
A key advantage of our model is its ability to maintain reconstruction accuracy under arbitrary rotations of the mask. While CNNs operating on flat maps naturally possess translation invariance, achieving rotation invariance for spherical maps presents unique challenges. 

\begin{figure*}
  \centering
  \includegraphics[width=0.8\textwidth]{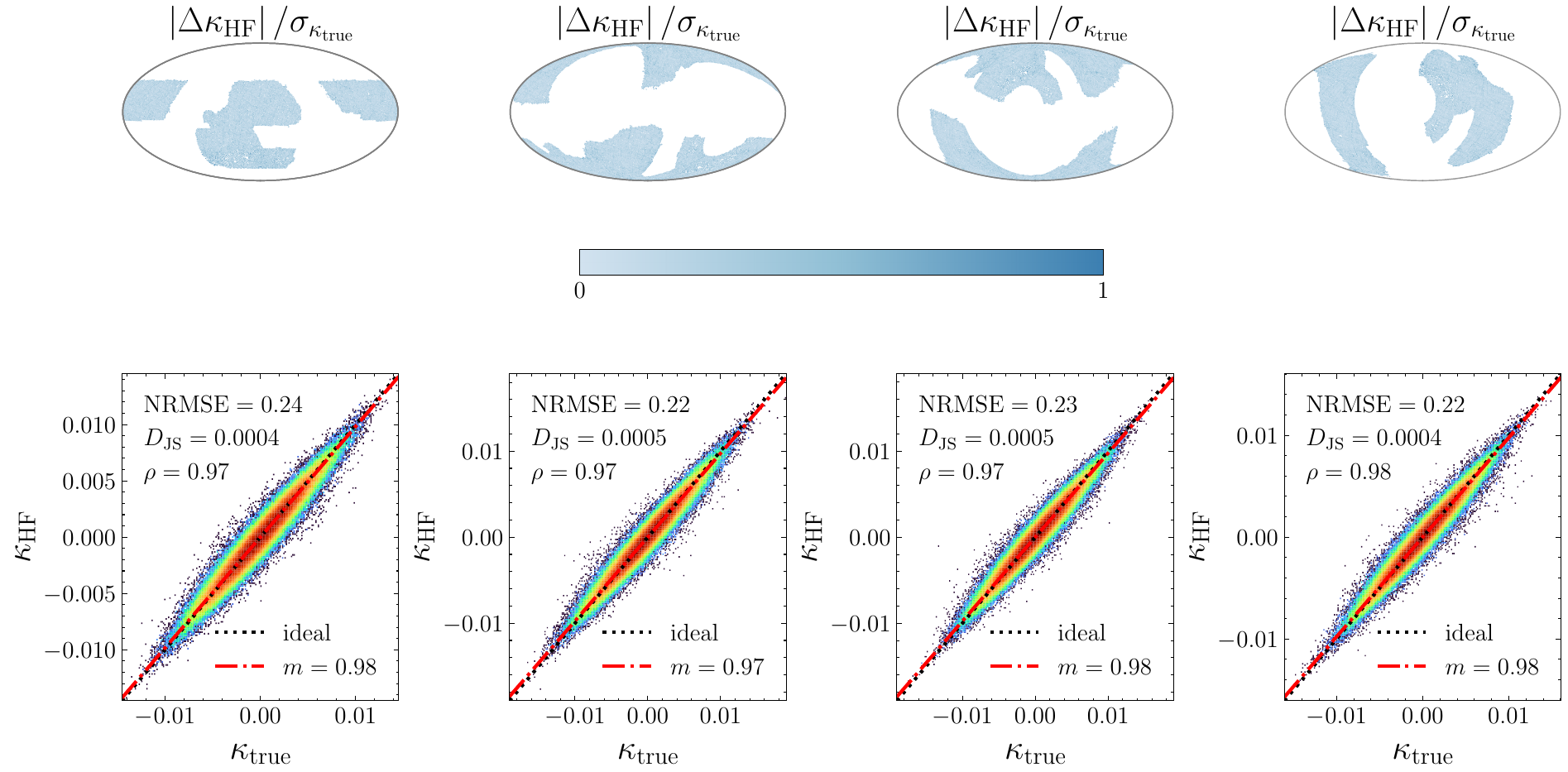}
  \caption{Demonstration of rotation invariance using the DECaLS mask at $N_{\mathrm{side}}=256$. The top row displays normalized residual maps while the bottom row shows pixel-by-pixel comparisons for different random rotations. Despite changes in mask geometry due to HEALPix pixelization, our model maintains consistent reconstruction quality across all rotations, demonstrating robust rotation invariance.}\label{fig:rotation}
\end{figure*}

To ensure rotation invariance, we incorporate random rotations during the training process by rotating the HEALPix maps on-the-fly. This data augmentation strategy proves essential, as vanilla position embeddings fail to maintain invariance under rotation due to their fixed relative position encoding. In contrast, our learnable position embeddings demonstrate remarkable adaptability to rotational transformations. As illustrated in Fig.~\ref{fig:rotation}, we evaluate this capability by applying random rotations to the DECaLS mask. The results show consistent reconstruction quality across multiple rotation angles, with nearly identical pixel-by-pixel comparisons between the original and rotated configurations, confirming the model's robust rotation invariance.

\subsection{Single model for different masks}\label{app:variable_sized_input}

Convolutional models such as DeepMass~\cite{jeffreyDeepLearningDark2020} require fixed-size inputs and therefore must be retrained for each survey footprint. Our transformer processes HEALPix data as variable-length sequences, so one network can handle many footprints without additional training. Because each mask reveals a different number of visible pixels, the input length changes from case to case, yet the model generalizes well to unseen masks. This design provides an efficient and adaptable solution for mass mapping across diverse survey geometries.

\begin{figure*}
  \centering
  \includegraphics[width=0.7\textwidth]{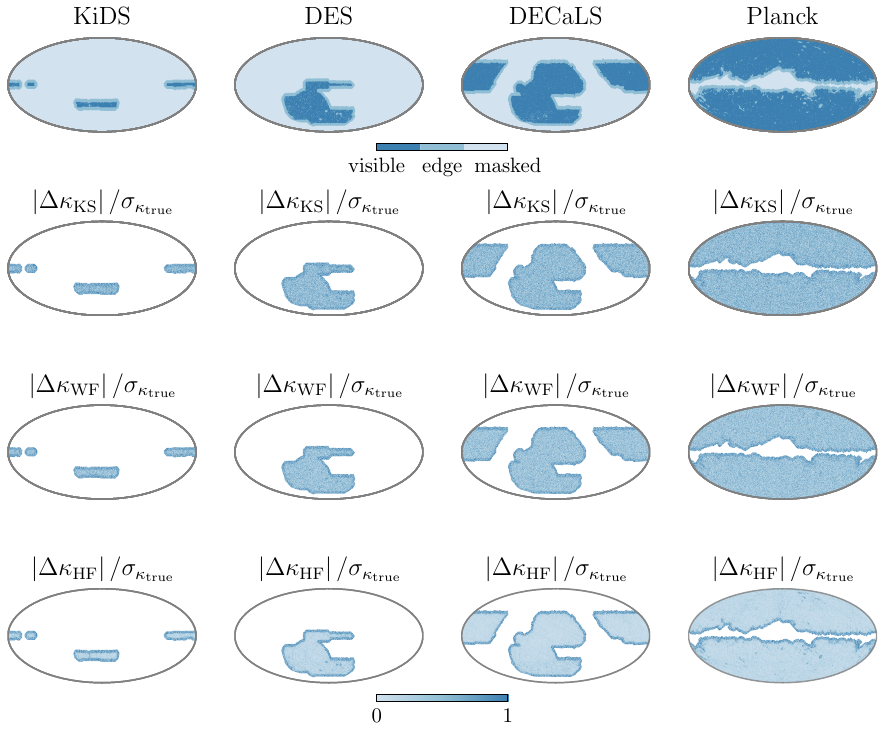}
  \caption{Comparison of normalized residual maps for different survey masks (KiDS, DES, DECaLS, and Planck) at $N_{\mathrm{side}}=256$. The first row shows the visible (dark blue), edge (light blue) and masked (grey) regions. The visible region contains observation data, while the edge region comprises pixels near survey boundaries and inner holes. The masked region represents pixels far from the survey boundary that are not relevant for reconstruction. The second through fourth rows show normalized residual maps for KS, WF, and HF reconstructions respectively. HF demonstrates superior performance with the lightest residuals, despite using a single model for all four masks. All methods show similar limitations in reconstructing outer edge, where limited information from visible regions constrains recovery capabilities.}\label{fig:all_masks}
\end{figure*}

To demonstrate this flexibility, we evaluated our model's performance across multiple survey configurations (KiDS, DES, DECaLS, and Planck masks) using a single trained model. As shown in Fig.~\ref{fig:all_masks}, our approach processes variable-sized inputs by retaining only visible and edge pixels, achieving superior reconstruction quality compared to traditional methods. While all methods, including ours, show reduced performance at outer edge due to limited information propagation from visible regions, HF consistently maintains better overall reconstruction accuracy across different mask configurations.

\bibliographystyle{apsrev4-2}
\bibliography{apssamp}

@article{Alonso_2019,
 author = {Alonso, David and Sanchez, Javier and Slosar, An{\v z}e},
 doi = {10.1093/mnras/stz093},
 issn = {1365-2966},
 journal = {Mon. Not. R. Astron. Soc.},
 month = {January},
 number = {3},
 pages = {4127--4151},
 publisher = {Oxford University Press (OUP)},
 title = {A Unified Pseudo-C{$\ell$} Framework},
 volume = {484},
 year = {2019}
}

@article{anselPyTorch2Faster2024,
 author = {Ansel, Jason and Yang, Edward and He, Horace and Gimelshein, Natalia and Jain, Animesh and Voznesensky, Michael and Bao, Bin and Bell, Peter and Berard, David and Burovski, Evgeni and Chauhan, Geeta and Chourdia, Anjali and Constable, Will and Desmaison, Alban and DeVito, Zachary and Ellison, Elias and Feng, Will and Gong, Jiong and Gschwind, Michael and Hirsh, Brian and Huang, Sherlock and Kalambarkar, Kshiteej and Kirsch, Laurent and Lazos, Michael and Lezcano, Mario and Liang, Yanbo and Liang, Jason and Lu, Yinghai and Luk, C. K. and Maher, Bert and Pan, Yunjie and Puhrsch, Christian and Reso, Matthias and Saroufim, Mark and Siraichi, Marcos Yukio and Suk, Helen and Zhang, Shunting and Suo, Michael and Tillet, Phil and Zhao, Xu and Wang, Eikan and Zhou, Keren and Zou, Richard and Wang, Xiaodong and Mathews, Ajit and Wen, William and Chanan, Gregory and Wu, Peng and Chintala, Soumith},
 doi = {10.1145/3620665.3640366},
 journal = {Proc. 29th ACM Int. Conf. Archit. Support Program. Lang. Oper. Syst. Vol. 2},
 month = {April},
 pages = {929--947},
 shorttitle = {PyTorch 2},
 title = {PyTorch 2: Faster Machine Learning Through Dynamic Python Bytecode Transformation and Graph Compilation},
 year = {2024}
}

@article{Bartelmann2001,
 author = {Bartelmann, Matthias and Schneider, Peter},
 doi = {10.1016/S0370-1573(00)00082-X},
 eprint = {astro-ph/9912508},
 issn = {03701573},
 journal = {Physics Reports},
 month = {January},
 number = {4-5},
 pages = {291--472},
 title = {Weak Gravitational Lensing},
 volume = {340},
 year = {2001}
}

@misc{bartelmannClusterMassEstimates1994,
 author = {Bartelmann, Matthias},
 doi = {10.48550/ARXIV.ASTRO-PH/9412051},
 publisher = {arXiv},
 title = {Cluster Mass Estimates from Weak Lensing},
 year = {1994}
}

@inproceedings{carlssonHEALSWINVisionTransformer2024,
 archiveprefix = {arXiv},
 author = {Carlsson, Oscar and Gerken, Jan E. and Linander, Hampus and Spie{\ss}, Heiner and Ohlsson, Fredrik and Petersson, Christoffer and Persson, Daniel},
 booktitle = {2024 CVPR},
 doi = {10.1109/CVPR52733.2024.00580},
 eprint = {2307.07313},
 month = {June},
 pages = {6067--6077},
 primaryclass = {cs},
 publisher = {\textbraceleft IEEE\textbraceright},
 shorttitle = {Heal-Swin},
 title = {HEAL-SWIN: A Vision Transformer on the Sphere},
 year = {2024}
}

@article{Castro-PhysRevD.72.023516,
 author = {Castro, P. G. and Heavens, A. F. and Kitching, T. D.},
 doi = {10.1103/PhysRevD.72.023516},
 issn = {1550-7998, 1550-2368},
 journal = {Phys. Rev. D},
 month = {July},
 number = {2},
 pages = {023516},
 publisher = {American Physical Society},
 title = {Weak Lensing Analysis in Three Dimensions},
 volume = {72},
 year = {2005}
}

@article{changDarkEnergySurvey2018,
 archiveprefix = {arXiv},
 author = {Chang, C. and Pujol, A. and Mawdsley, B. and Bacon, D. and {Elvin-Poole}, J. and Melchior, P. and Kov{\'a}cs, A. and Jain, B. and Leistedt, B. and Giannantonio, T. and Alarcon, A. and Baxter, E. and Bechtol, K. and Becker, M. R. and {Benoit-L{\'e}vy}, A. and Bernstein, G. M. and Bonnett, C. and Busha, M. T. and Rosell, A. Carnero and Castander, F. J. and Cawthon, R. and {da Costa}, L. N. and Davis, C. and De Vicente, J. and DeRose, J. and {Drlica- Wagner}, A. and Fosalba, P. and Gatti, M. and Gaztanaga, E. and Gruen, D. and Gschwend, J. and Hartley, W. G. and Hoyle, B. and Huff, E. M. and Jarvis, M. and Jeffrey, N. and Kacprzak, T. and Lin, H. and MacCrann, N. and Maia, M. A. G. and Ogando, R. L. C. and Prat, J. and Rau, M. M. and Rollins, R. P. and Roodman, A. and Rozo, E. and Rykoff, E. S. and Samuroff, S. and S{\'a}nchez, C. and {Sevilla-Noarbe}, I. and Sheldon, E. and Troxel, M. A. and Varga, T. N. and Vielzeuf, P. and Vikram, V. and Wechsler, R. H. and Zuntz, J. and Abbott, T. M. C. and Abdalla, F. B. and Allam, S. and Annis, J. and Bertin, E. and Brooks, D. and {Buckley-Geer}, E. and Burke, D. L. and Kind, M. Carrasco and Carretero, J. and Crocce, M. and Cunha, C. E. and D'Andrea, C. B. and Desai, S. and Diehl, H. T. and Dietrich, J. P. and Doel, P. and Estrada, J. and Neto, A. Fausti and Fernandez, E. and Flaugher, B. and Frieman, J. and {Garc{\'i}a-Bellido}, J. and Gruendl, R. A. and Gutierrez, G. and Honscheid, K. and James, D. J. and Jeltema, T. and Johnson, M. W. G. and Johnson, M. D. and Kent, S. and Kirk, D. and Krause, E. and Kuehn, K. and Kuhlmann, S. and Lahav, O. and Li, T. S. and Lima, M. and March, M. and Martini, P. and Menanteau, F. and Miquel, R. and Mohr, J. J. and Neilsen, E. and Nichol, R. C. and Petravick, D. and Plazas, A. A. and Romer, A. K. and Sako, M. and Sanchez, E. and Scarpine, V. and Schubnell, M. and Smith, M. and Smith, R. C. and {Soares-Santos}, M. and Sobreira, F. and Suchyta, E. and Tarle, G. and Thomas, D. and Tucker, D. L. and Walker, A. R. and Wester, W. and Zhang, Y.},
 doi = {10.1093/mnras/stx3363},
 eprint = {1708.01535},
 issn = {0035-8711, 1365-2966},
 journal = {Mon. Not. R. Astron. Soc.},
 month = {April},
 number = {3},
 pages = {3165--3190},
 primaryclass = {astro-ph.CO},
 shorttitle = {Dark Energy Survey Year 1 Results},
 title = {Dark Energy Survey Year 1 Results: Curved-Sky Weak Lensing Mass Map},
 volume = {475},
 year = {2018}
}

@article{changEffectiveNumberDensity2013,
 author = {Chang, C. and Jarvis, M. and Jain, B. and Kahn, S. M. and Kirkby, D. and Connolly, A. and Krughoff, S. and Peng, E.- H. and Peterson, J. R.},
 doi = {10.1093/mnras/stt1156},
 issn = {0035-8711, 1365-2966},
 journal = {Mon. Not. R. Astron. Soc.},
 month = {September},
 number = {3},
 pages = {2121--2135},
 title = {The Effective Number Density of Galaxies for Weak Lensing Measurements in the LSST Project},
 volume = {434},
 year = {2013}
}

@article{chisariCoreCosmologyLibrary2019,
 archiveprefix = {arXiv},
 author = {Chisari, Nora Elisa and Alonso, David and Krause, Elisabeth and Leonard, C. Danielle and Bull, Philip and Neveu, J{\'e}r{\'e}my and Villarreal, Antonio and Singh, Sukhdeep and McClintock, Thomas and Ellison, John and Du, Zilong and Zuntz, Joe and Mead, Alexander and Joudaki, Shahab and Lorenz, Christiane S. and Tr{\"o}ster, Tilman and Sanchez, Javier and Lanusse, Francois and Ishak, Mustapha and Hlozek, Ren{\'e}e and Blazek, Jonathan and Campagne, Jean-Eric and Almoubayyed, Husni and Eifler, Tim and Kirby, Matthew and Kirkby, David and Plaszczynski, St{\'e}phane and Slosar, An{\v z}e and Vrastil, Michal and Wagoner, Erika L. and {(LSST Dark Energy Science Collaboration)}},
 doi = {10.3847/1538-4365/ab1658},
 eprint = {1812.05995},
 issn = {0067-0049, 1538-4365},
 journal = {ApJS},
 month = {May},
 number = {1},
 pages = {2},
 primaryclass = {astro-ph},
 shorttitle = {Core Cosmology Library},
 title = {Core Cosmology Library: Precision Cosmological Predictions for LSST},
 volume = {242},
 year = {2019}
}

@inproceedings{cipollaMultitaskLearningUsing2018,
 address = {Salt Lake City, UT, USA},
 author = {Cipolla, Roberto and Gal, Yarin and Kendall, Alex},
 booktitle = {2018 CVPR},
 doi = {10.1109/CVPR.2018.00781},
 isbn = {978-1-5386-6420-9},
 month = {June},
 pages = {7482--7491},
 publisher = {Computer Vision Foundation / \textbraceleft IEEE\textbraceright{} Computer Society},
 title = {Multi-Task Learning Using Uncertainty to Weigh Losses for Scene Geometry and Semantics},
 year = {2018}
}

@misc{collaborationIntroductionChinaSpace2025,
 archiveprefix = {arXiv},
 author = {Collaboration, Csst and Gong, Yan and Miao, Haitao and Zhan, Hu and Li, Zhao-Yu and Shangguan, Jinyi and Li, Haining and Liu, Chao and Chen, Xuefei and Yuan, Haibo and Zhou, Jilin and Liu, Hui-Gen and Yu, Cong and Ji, Jianghui and Qi, Zhaoxiang and Liu, Jiacheng and Dai, Zigao and Wang, Xiaofeng and Zheng, Zhenya and Hao, Lei and Dou, Jiangpei and Ao, Yiping and Lin, Zhenhui and Zhang, Kun and Wang, Wei and Sun, Guotong and Li, Ran and Li, Guoliang and Xu, Youhua and Li, Xinfeng and Li, Shengyang and Wu, Peng and Zhang, Jiuxing and Wang, Bo and Bai, Jinming and Cai, Yi-Fu and Cai, Zheng and Chan, Kwan Chuen and Chang, Jin and Chen, Xiaodian and Chen, Xuelei and Chen, Yuqin and Chen, Yun and Cui, Wei and Du, Pu and Duan, Wenying and Fan, Junhui and Fan, LuLu and Fan, Zhou and Fan, Zuhui and Fang, Taotao and Fu, Jianning and Fu, Liping and Fu, Zhensen and Gao, Jian and Gu, Shenghong and Gu, Yidong and Guo, Qi and Han, Zhanwen and Huang, Zhiqi and Ho, Luis C. and Jiang, Linhua and Jing, Yipeng and Kang, Xi and Kong, Xu and Li, Chengyuan and Li, Di and Li, Jing and Li, Nan and Li, Yang A. and Liao, Shilong and Lin, Weipeng and Liu, Fengshan and Liu, Jifeng and Liu, Xiangkun and Mao, Ruiqing and Mao, Shude and Meng, Xianmin and Pang, Xiaoying and Peng, Xiyan and Peng, Yingjie and Shan, Huanyuan and Shen, Juntai and Shen, Shiyin and Shen, Zhiqiang and Shi, Sheng-Cai and Shi, Yong and Tan, Siyuan and Tian, Hao and Wang, Jianmin and Wang, Jun-Xian and Wang, Xin and Wang, Yuting and Wu, Hong and Wu, Jingwen and Wu, Xuebing and Xu, Chun and Xue, Xiang-Xiang and Xue, Yongquan and Yang, Ji and Yang, Xiaohu and Yao, Qijun and Yuan, Fangting and Yuan, Zhen and Zhang, Jun and Zhang, Wei and Zhang, Xin and Zhao, Gang and Zhao, Gongbo and Zhong, Hongen and Zhong, Jing and Zhou, Liyong and Zu, Ying},
 doi = {10.48550/arXiv.2507.04618},
 eprint = {2507.04618},
 month = {July},
 number = {arXiv:2507.04618},
 primaryclass = {astro-ph},
 publisher = {arXiv},
 title = {Introduction to the China Space Station Telescope (CSST)},
 year = {2025}
}

@misc{collaborationLargeSynopticSurvey2012,
 archiveprefix = {arXiv},
 author = {Collaboration, LSST Dark Energy Science},
 doi = {10.48550/arXiv.1211.0310},
 eprint = {1211.0310},
 month = {November},
 number = {arXiv:1211.0310},
 primaryclass = {astro-ph},
 publisher = {arXiv},
 shorttitle = {Large Synoptic Survey Telescope},
 title = {Large Synoptic Survey Telescope: Dark Energy Science Collaboration},
 year = {2012}
}

@inproceedings{defazioRoadLessScheduled2024,
 archiveprefix = {arXiv},
 author = {Defazio, Aaron and Yang, Xingyu and Khaled, Ahmed and Mishchenko, Konstantin and Mehta, Harsh and Cutkosky, Ashok},
 booktitle = {NeurIPS 2024},
 doi = {10.48550/arXiv.2405.15682},
 eprint = {2405.15682},
 primaryclass = {cs},
 publisher = {arXiv},
 title = {The Road Less Scheduled},
 year = {2024}
}

@inproceedings{devlinBERTPretrainingDeep2019,
 address = {Minneapolis, Minnesota},
 author = {Devlin, Jacob and Chang, Ming{\textbraceleft}-{\textbraceright}Wei and Lee, Kenton and Toutanova, Kristina},
 booktitle = {2019 NAACL},
 doi = {10.18653/V1/N19-1423},
 pages = {4171--4186},
 publisher = {Association for Computational Linguistics},
 title = {\textbraceleft BERT:\textbraceright{} Pre-Training of Deep Bidirectional Transformers for Language Understanding},
 year = {2019}
}

@article{deyOverviewDESILegacy2019,
 author = {Dey, Arjun and Schlegel, David J. and Lang, Dustin and Blum, Robert and Burleigh, Kaylan and Fan, Xiaohui and Findlay, Joseph R. and Finkbeiner, Doug and Herrera, David and Juneau, St{\'e}phanie and Landriau, Martin and Levi, Michael and McGreer, Ian and Meisner, Aaron and Myers, Adam D. and Moustakas, John and Nugent, Peter and Patej, Anna and Schlafly, Edward F. and Walker, Alistair R. and Valdes, Francisco and Weaver, Benjamin A. and Y{\`e}che, Christophe and Zou, Hu and Zhou, Xu and Abareshi, Behzad and Abbott, T. M. C. and Abolfathi, Bela and Aguilera, C. and Alam, Shadab and Allen, Lori and Alvarez, A. and Annis, James and Ansarinejad, Behzad and Aubert, Marie and Beechert, Jacqueline and Bell, Eric F. and BenZvi, Segev Y. and Beutler, Florian and Bielby, Richard M. and Bolton, Adam S. and Brice{\~n}o, C{\'e}sar and {Buckley-Geer}, Elizabeth J. and Butler, Karen and Calamida, Annalisa and Carlberg, Raymond G. and Carter, Paul and Casas, Ricard and Castander, Francisco J. and Choi, Yumi and Comparat, Johan and Cukanovaite, Elena and Delubac, Timoth{\'e}e and DeVries, Kaitlin and Dey, Sharmila and Dhungana, Govinda and Dickinson, Mark and Ding, Zhejie and Donaldson, John B. and Duan, Yutong and Duckworth, Christopher J. and Eftekharzadeh, Sarah and Eisenstein, Daniel J. and Etourneau, Thomas and Fagrelius, Parker A. and Farihi, Jay and Fitzpatrick, Mike and {Font-Ribera}, Andreu and Fulmer, Leah and G{\"a}nsicke, Boris T. and Gaztanaga, Enrique and George, Koshy and Gerdes, David W. and A Gontcho, Satya Gontcho and Gorgoni, Claudio and Green, Gregory and Guy, Julien and Harmer, Diane and Hernandez, M. and Honscheid, Klaus and Huang, Lijuan (Wendy) and James, David J. and Jannuzi, Buell T. and Jiang, Linhua and Joyce, Richard and Karcher, Armin and Karkar, Sonia and Kehoe, Robert and Kneib, Jean-Paul and {Kueter-Young}, Andrea and Lan, Ting-Wen and Lauer, Tod R. and Guillou, Laurent Le and Van Suu, Auguste Le and Lee, Jae Hyeon and Lesser, Michael and Levasseur, Laurence Perreault and Li, Ting S. and Mann, Justin L. and Marshall, Robert and {Mart{\'i}nez-V{\'a}zquez}, C. E. and Martini, Paul and Du Mas Des Bourboux, H{\'e}lion and McManus, Sean and Meier, Tobias Gabriel and M{\'e}nard, Brice and Metcalfe, Nigel and {Mu{\~n}oz-Guti{\'e}rrez}, Andrea and Najita, Joan and Napier, Kevin and Narayan, Gautham and Newman, Jeffrey A. and Nie, Jundan and Nord, Brian and Norman, Dara J. and Olsen, Knut A. G. and Paat, Anthony and {Palanque-Delabrouille}, Nathalie and Peng, Xiyan and Poppett, Claire L. and Poremba, Megan R. and Prakash, Abhishek and Rabinowitz, David and Raichoor, Anand and Rezaie, Mehdi and Robertson, A. N. and Roe, Natalie A. and Ross, Ashley J. and Ross, Nicholas P. and Rudnick, Gregory and Gaines, Sasha and Saha, Abhijit and S{\'a}nchez, F. Javier and Savary, Elodie and Schweiker, Heidi and Scott, Adam and Seo, Hee-Jong and Shan, Huanyuan and Silva, David R. and Slepian, Zachary and Soto, Christian and Sprayberry, David and Staten, Ryan and Stillman, Coley M. and Stupak, Robert J. and Summers, David L. and Tie, Suk Sien and Tirado, H. and {Vargas-Maga{\~n}a}, Mariana and Vivas, A. Katherina and Wechsler, Risa H. and Williams, Doug and Yang, Jinyi and Yang, Qian and Yapici, Tolga and Zaritsky, Dennis and Zenteno, A. and Zhang, Kai and Zhang, Tianmeng and Zhou, Rongpu and Zhou, Zhimin},
 doi = {10.3847/1538-3881/ab089d},
 issn = {0004-6256, 1538-3881},
 journal = {Astron. J.},
 month = {May},
 number = {5},
 pages = {168},
 publisher = {American Astronomical Society},
 title = {Overview of the DESI Legacy Imaging Surveys},
 volume = {157},
 year = {2019}
}

@misc{dosovitskiyImageWorth16x162020,
 archiveprefix = {arXiv},
 author = {Dosovitskiy, Alexey and Beyer, Lucas and Kolesnikov, Alexander and Weissenborn, Dirk and Zhai, Xiaohua and Unterthiner, Thomas and Dehghani, Mostafa and Minderer, Matthias and Heigold, Georg and Gelly, Sylvain and Uszkoreit, Jakob and Houlsby, Neil},
 eprint = {2010.11929},
 publisher = {arXiv},
 title = {An Image Is Worth 16x16 Words: Transformers for Image Recognition at Scale},
 year = {2021}
}

@article{fiedorowiczKaRMMaKappaReconstruction2022a,
 archiveprefix = {arXiv},
 author = {Fiedorowicz, Pier and Rozo, Eduardo and Boruah, Supranta S. and Chang, Chihway and Gatti, Marco},
 doi = {10.1093/mnras/stac468},
 eprint = {2105.14699},
 issn = {0035-8711, 1365-2966},
 journal = {Mon. Not. R. Astron. Soc.},
 month = {March},
 number = {1},
 pages = {73--85},
 primaryclass = {astro-ph},
 title = {KaRMMa -- Kappa Reconstruction for Mass Mapping},
 volume = {512},
 year = {2022}
}

@misc{gehringConvolutionalSequenceSequence2017,
 archiveprefix = {arXiv},
 author = {Gehring, Jonas and Auli, Michael and Grangier, David and Yarats, Denis and Dauphin, Yann N.},
 doi = {10.48550/arXiv.1705.03122},
 eprint = {1705.03122},
 month = {July},
 number = {arXiv:1705.03122},
 primaryclass = {cs},
 publisher = {arXiv},
 title = {Convolutional Sequence to Sequence Learning},
 year = {2017}
}

@article{GORSKI2005,
 author = {G{\'o}rski, K. M. and Hivon, E. and Banday, A. J. and Wandelt, B. D. and Hansen, F. K. and Reinecke, M. and Bartelmann, M.},
 doi = {10.1086/427976},
 eprint = {arXiv:astro-ph/0409513},
 issn = {0004-637X, 1538-4357},
 journal = {Astrophys. J.},
 month = {April},
 number = {2},
 pages = {759--771},
 shorttitle = {HEALPix},
 title = {HEALPix: A Framework for High-Resolution Discretization and Fast Analysis of Data Distributed on the Sphere},
 volume = {622},
 year = {2005}
}

@article{grewalComparingMassMapping2024,
 archiveprefix = {arXiv},
 author = {Grewal, Nisha and Zuntz, Joe and Tr{\"o}ster, Tilman},
 doi = {10.33232/001c.120394},
 eprint = {2402.13912},
 issn = {2565-6120},
 journal = {Open J. Astrophys.},
 month = {June},
 primaryclass = {astro-ph},
 title = {Comparing Mass Mapping Reconstruction Methods with Minkowski Functionals},
 volume = {7},
 year = {2024}
}

@article{harrisArrayProgrammingNumPy2020,
 author = {Harris, Charles R. and Millman, K. Jarrod and Van Der Walt, St{\'e}fan J. and Gommers, Ralf and Virtanen, Pauli and Cournapeau, David and Wieser, Eric and Taylor, Julian and Berg, Sebastian and Smith, Nathaniel J. and Kern, Robert and Picus, Matti and Hoyer, Stephan and Van Kerkwijk, Marten H. and Brett, Matthew and Haldane, Allan and Del R{\'i}o, Jaime Fern{\'a}ndez and Wiebe, Mark and Peterson, Pearu and {G{\'e}rard-Marchant}, Pierre and Sheppard, Kevin and Reddy, Tyler and Weckesser, Warren and Abbasi, Hameer and Gohlke, Christoph and Oliphant, Travis E.},
 doi = {10.1038/s41586-020-2649-2},
 issn = {0028-0836, 1476-4687},
 journal = {Nature},
 month = {September},
 number = {7825},
 pages = {357--362},
 title = {Array Programming with NumPy},
 volume = {585},
 year = {2020}
}

@inproceedings{heMaskedAutoencodersAre,
 archiveprefix = {arXiv},
 author = {He, Kaiming and Chen, Xinlei and Xie, Saining and Li, Yanghao and Dollar, Piotr and Girshick, Ross},
 booktitle = {2022 CVPR},
 doi = {10.1109/CVPR52688.2022.01553},
 eprint = {2111.06377v1},
 month = {June},
 pages = {15979--15988},
 publisher = {\textbraceleft IEEE\textbraceright},
 title = {Masked Autoencoders Are Scalable Vision Learners},
 year = {2022}
}

@article{heymansKiDS1000CosmologyMultiprobe2021,
 archiveprefix = {arXiv},
 author = {Heymans, Catherine and Tr{\"o}ster, Tilman and Asgari, Marika and Blake, Chris and Hildebrandt, Hendrik and Joachimi, Benjamin and Kuijken, Konrad and Lin, Chieh-An and S{\'a}nchez, Ariel G. and van den Busch, Jan Luca and Wright, Angus H. and Amon, Alexandra and Bilicki, Maciej and de Jong, Jelte and Crocce, Martin and Dvornik, Andrej and Erben, Thomas and Fortuna, Maria Cristina and Getman, Fedor and Giblin, Benjamin and Glazebrook, Karl and Hoekstra, Henk and Joudaki, Shahab and Kannawadi, Arun and K{\"o}hlinger, Fabian and Lidman, Chris and Miller, Lance and Napolitano, Nicola R. and Parkinson, David and Schneider, Peter and Shan, HuanYuan and Valentijn, Edwin and Kleijn, Gijs Verdoes and Wolf, Christian},
 doi = {10.1051/0004-6361/202039063},
 eprint = {2007.15632},
 issn = {0004-6361, 1432-0746},
 journal = {Astron. Astrophys.},
 month = {February},
 pages = {A140},
 primaryclass = {astro-ph},
 shorttitle = {KiDS-1000 Cosmology},
 title = {KiDS-1000 Cosmology: Multi-Probe Weak Gravitational Lensing and Spectroscopic Galaxy Clustering Constraints},
 volume = {646},
 year = {2021}
}

@article{Hildebrandt2017,
 archiveprefix = {arXiv},
 author = {Hildebrandt, H. and Viola, M. and Heymans, C. and Joudaki, S. and Kuijken, K. and Blake, C. and Erben, T. and Joachimi, B. and Klaes, D. and Miller, L. and Morrison, C. B. and Nakajima, R. and Verdoes Kleijn, G. and Amon, A. and Choi, A. and Covone, G. and {de Jong}, J. T. A. and Dvornik, A. and Fenech Conti, I. and Grado, A. and {Harnois-D{\'e}raps}, J. and Herbonnet, R. and Hoekstra, H. and K{\"o}hlinger, F. and McFarland, J. and Mead, A. and Merten, J. and Napolitano, N. and Peacock, J. A. and Radovich, M. and Schneider, P. and Simon, P. and Valentijn, E. A. and {van den Busch}, J. L. and {van Uitert}, E. and Van Waerbeke, L.},
 doi = {10.1093/mnras/stw2805},
 eprint = {1606.05338},
 issn = {0035-8711, 1365-2966},
 journal = {Mon. Not. R. Astron. Soc.},
 month = {February},
 number = {2},
 pages = {1454--1498},
 shorttitle = {KiDS-450},
 title = {KiDS-450: Cosmological Parameter Constraints from Tomographic Weak Gravitational Lensing},
 volume = {465},
 year = {2017}
}

@article{huLoRALowRankAdaptation2021,
 archiveprefix = {arXiv},
 author = {Hu, J. E. and Shen, Yelong and Wallis, Phillip and {Allen-Zhu}, Zeyuan and Li, Yuanzhi and Wang, Shean and Chen, Weizhu},
 doi = {10.48550/arXiv.2106.09685},
 eprint = {2106.09685},
 journal = {ArXiv},
 primaryclass = {cs},
 shorttitle = {LoRA},
 title = {LoRA: Low-Rank Adaptation of Large Language Models},
 volume = {abs/2106.09685},
 year = {2021}
}

@article{jeffreyDarkEnergySurvey2021,
 archiveprefix = {arXiv},
 author = {Jeffrey, N and Gatti, M and Chang, {\dag} C and Whiteway, L and Demirbozan, U and Kovacs, A and Pollina, G and Bacon, D and Hamaus, N and Kacprzak, T and Lahav, O and Lanusse, F and Mawdsley, B and Nadathur, S and Starck, J L and Vielzeuf, P and Zeurcher, D and Alarcon, A and Amon, A and Bechtol, K and Bernstein, G M and Campos, A and Rosell, A Carnero and Carrasco, M and Cawthon, R and Chen, R and Choi, A and Cordero, J and Davis, C and Derose, J and Doux, C and {Drlica-Wagner}, A and Eckert, K and Elsner, F and {Elvin-Poole}, J and Everett, S and Fert{\'e}, A and Giannini, G and Gruen, D and Gruendl, R A and Harrison, I and Hartley, W G and Herner, K and Huff, E M and Huterer, D and Kuropatkin, N and Jarvis, M and Leget, P F and Maccrann, N and Mccullough, J and Muir, J and Myles, J and {Navarro-Alsina}, A and Pandey, S and Prat, J and Raveri, M and Rollins, R P and Ross, A J and Rykoff, E S and S{\'a}nchez, C and Secco, L F and {Sevilla-Noarbe}, I and Sheldon, E and Shin, T and Troxel, M A and Tutusaus, I and Varga, T N and Yanny, B and Yin, B and Zhang, Y and Zuntz, J and Abbott, T M C and Aguena, M and Allam, S and {Andrade-Oliveira}, F and Becker, M R and Bertin, E and Bhargava, S and Brooks, D and Burke, D L and Carretero, J},
 eprint = {2105.13539v2},
 issn = {2019-0489},
 journal = {J. Gschwend},
 pages = {9},
 title = {Dark Energy Survey Year 3 Results: Curved-Sky Weak Lensing Mass Map Reconstruction},
 volume = {18},
 year = {2021}
}

@article{jeffreyDeepLearningDark2020,
 archiveprefix = {arXiv},
 author = {Jeffrey, Niall and Lanusse, Fran{\c c}ois and Lahav, Ofer and Starck, Jean Luc},
 doi = {10.1093/MNRAS/STAA127},
 eprint = {1908.00543},
 issn = {13652966},
 journal = {Mon. Not. R. Astron. Soc.},
 number = {4},
 pages = {5023--5029},
 title = {Deep Learning Dark Matter Map Reconstructions from Des SV Weak Lensing Data},
 volume = {492},
 year = {2020}
}

@article{jeffreyFastSamplingWiener2018,
 archiveprefix = {arXiv},
 author = {Jeffrey, N. and Heavens, A. F. and Fortio, P. D.},
 doi = {10.1016/j.ascom.2018.10.001},
 eprint = {1810.02821},
 issn = {22131337},
 journal = {Astron. Comput.},
 month = {October},
 pages = {230--237},
 primaryclass = {astro-ph.IM},
 title = {Fast Sampling from Wiener Posteriors for Image Data with Dataflow Engines},
 volume = {25},
 year = {2018}
}

@article{kaiserMappingDarkMatter1993,
 author = {Kaiser, Nick and Squires, Gordon},
 doi = {10.1086/172297},
 issn = {0004-637X, 1538-4357},
 journal = {Astrophys. J.},
 month = {February},
 pages = {441},
 title = {Mapping the Dark Matter with Weak Gravitational Lensing},
 volume = {404},
 year = {1993}
}

@inproceedings{kingma2014adam,
 author = {Kingma, Diederik P. and Ba, Jimmy},
 booktitle = {Proc. 3rd Int. Conf. Learn. Represent.},
 title = {Adam: \textbraceleft a\textbraceright{} Method for Stochastic Optimization},
 year = {2015}
}

@article{lahavWienerReconstructionAllSky1994,
 author = {Lahav, O. and Fisher, K. B. and Hoffman, Y. and Scharf, C. A. and Zaroubi, S.},
 doi = {10.1086/187244},
 issn = {0004-637X, 1538-4357},
 journal = {Astrophys. J.},
 month = {March},
 pages = {L93},
 title = {Wiener Reconstruction of All-Sky Galaxy Surveys in Spherical Harmonics},
 volume = {423},
 year = {1994}
}

@article{leonardGLIMPSEAccurate3D2014,
 author = {Leonard, Adrienne and Lanusse, Fran{\c c}ois and Starck, Jean-Luc},
 doi = {10.1093/mnras/stu273},
 issn = {0035-8711, 1365-2966},
 journal = {Mon. Not. R. Astron. Soc.},
 month = {May},
 number = {2},
 pages = {1281--1294},
 shorttitle = {Glimpse},
 title = {GLIMPSE: Accurate 3D Weak Lensing Reconstructions Using Sparsity},
 volume = {440},
 year = {2014}
}

@article{mandelbaumWeakLensingPrecision2018,
 archiveprefix = {arXiv},
 author = {Mandelbaum, Rachel},
 doi = {10.1146/annurev-astro-081817-051928},
 eprint = {1710.03235},
 issn = {0066-4146, 1545-4282},
 journal = {Annu. Rev. Astron. Astrophys.},
 month = {September},
 number = {1},
 pages = {393--433},
 primaryclass = {astro-ph.CO},
 title = {Weak Lensing for Precision Cosmology},
 volume = {56},
 year = {2018}
}

@article{masseyDarkMatterMaps2007,
 author = {Massey, Richard and Rhodes, Jason and Ellis, Richard and Scoville, Nick and Leauthaud, Alexie and Finoguenov, Alexis and Capak, Peter and Bacon, David and Aussel, Herv{\'e} and Kneib, Jean-Paul and Koekemoer, Anton and McCracken, Henry and Mobasher, Bahram and Pires, Sandrine and Refregier, Alexandre and Sasaki, Shunji and Starck, Jean-Luc and Taniguchi, Yoshi and Taylor, Andy and Taylor, James},
 doi = {10.1038/nature05497},
 issn = {0028-0836, 1476-4687},
 journal = {Nature},
 month = {January},
 number = {7125},
 pages = {286--290},
 title = {Dark Matter Maps Reveal Cosmic Scaffolding},
 volume = {445},
 year = {2007}
}

@article{mawdsleyDarkEnergySurvey2020,
 author = {Mawdsley, B and Bacon, D and Chang, C and Melchior, P and Rozo, E and Seitz, S and Jeffrey, N and Gatti, M and Gaztanaga, E and Gruen, D and Hartley, W G and Hoyle, B and Samuroff, S and Sheldon, E and Troxel, M A and Zuntz, J and Abbott, T M C and Annis, J and Bertin, E and Bridle, S L and Brooks, D and {Buckley-Geer}, E and Burke, D L and Carnero~Rosell, A and Carrasco~Kind, M and Carretero, J and {da~Costa}, L N and De~Vicente, J and Desai, S and Diehl, H T and Doel, P and Evrard, A E and Flaugher, B and Fosalba, P and Frieman, J and {Garc{\'i}a-Bellido}, J and Gerdes, D W and Gruendl, R A and Gschwend, J and Gutierrez, G and Hollowood, D L and Honscheid, K and James, D J and Jarvis, M and Jeltema, T and Kuehn, K and Kuropatkin, N and Lima, M and Maia, M A G and Marshall, J L and Miquel, R and Plazas, A A and Roodman, A and Sanchez, E and Scarpine, V and Serrano, S and {Sevilla-Noarbe}, I and Smith, M and Smith, R C and Sobreira, F and Suchyta, E and Swanson, M E C and Tarle, G and Tucker, D L and Vikram, V and Walker, A R and {(DES Collaboration)}},
 doi = {10.1093/mnras/staa565},
 issn = {0035-8711, 1365-2966},
 journal = {Mon. Not. R. Astron. Soc.},
 month = {April},
 number = {4},
 pages = {5662--5679},
 shorttitle = {Dark Energy Survey Year 1 Results},
 title = {Dark Energy Survey Year 1 Results: Wide-Field Mass Maps via Forward Fitting in Harmonic Space},
 volume = {493},
 year = {2020}
}

@article{nguyenImageWorthMore2025,
 archiveprefix = {arXiv},
 author = {Nguyen, Duy-Kien and Assran, Mahmoud and Jain, Unnat and Oswald, Martin R. and Snoek, Cees G. M. and Chen, Xinlei},
 doi = {10.48550/ARXIV.2406.09415},
 eprint = {2406.09415},
 journal = {Corr},
 primaryclass = {cs},
 shorttitle = {An Image Is Worth More than 16x16 Patches},
 title = {An Image Is Worth More Than 16x16 Patches: Exploring Transformers on Individual Pixels},
 volume = {abs/2406.9415},
 year = {2024}
}

@article{perraudinDeepSphereEfficientSpherical2019,
 archiveprefix = {arXiv},
 author = {Perraudin, N. and Defferrard, M. and Kacprzak, T. and Sgier, R.},
 doi = {10.1016/j.ascom.2019.03.004},
 eprint = {1810.12186},
 issn = {22131337},
 journal = {Astron. Comput.},
 pages = {130--146},
 shorttitle = {DeepSphere},
 title = {DeepSphere: Efficient Spherical Convolutional Neural Network with HEALPix Sampling for Cosmological Applications},
 volume = {27},
 year = {2019}
}

@article{piresEuclidReconstructionWeaklensing2020,
 author = {Pires, S. and Vandenbussche, V. and Kansal, V. and Bender, R. and Blot, L. and Bonino, D. and Boucaud, A. and Brinchmann, J. and Capobianco, V. and Carretero, J. and Castellano, M. and Cavuoti, S. and Cl{\'e}dassou, R. and Congedo, G. and Conversi, L. and Corcione, L. and Dubath, F. and Fosalba, P. and Frailis, M. and Franceschi, E. and Fumana, M. and Grupp, F. and Hormuth, F. and Kermiche, S. and Knabenhans, M. and Kohley, R. and Kubik, B. and Kunz, M. and Ligori, S. and Lilje, P. B. and Lloro, I. and Maiorano, E. and Marggraf, O. and Massey, R. and Meylan, G. and Padilla, C. and Paltani, S. and Pasian, F. and Poncet, M. and Potter, D. and Raison, F. and Rhodes, J. and Roncarelli, M. and Saglia, R. and Schneider, P. and Secroun, A. and Serrano, S. and Stadel, J. and Tallada Cresp{\'i}, P. and Tereno, I. and {Toledo-Moreo}, R. and Wang, Y.},
 doi = {10.1051/0004-6361/201936865},
 issn = {0004-6361, 1432-0746},
 journal = {A\&A},
 month = {June},
 pages = {A141},
 shorttitle = {{\emph{Euclid}}},
 title = {{\emph{Euclid}} : Reconstruction of Weak-Lensing Mass Maps for Non-Gaussianity Studies},
 volume = {638},
 year = {2020}
}

@article{planckcollaborationPlanck2018Results2018,
 archiveprefix = {arXiv},
 author = {{Planck Collaboration} and Akrami, Y. and Arroja, F. and Ashdown, M. and Aumont, J. and Baccigalupi, C. and Ballardini, M. and Banday, A. J. and Barreiro, R. B. and Bartolo, N. and Basak, S. and Battye, R. and Benabed, K. and Bernard, J. -P. and Bersanelli, M. and Bielewicz, P. and Bock, J. J. and Bond, J. R. and Borrill, J. and Bouchet, F. R. and Boulanger, F. and Bucher, M. and Burigana, C. and Butler, R. C. and Calabrese, E. and Cardoso, J. -F. and Carron, J. and Casaponsa, B. and Challinor, A. and Chiang, H. C. and Colombo, L. P. L. and Combet, C. and Contreras, D. and Crill, B. P. and Cuttaia, F. and {de Bernardis}, P. and {de Zotti}, G. and Delabrouille, J. and Delouis, J. -M. and D{\'e}sert, F. -X. and Di Valentino, E. and Dickinson, C. and Diego, J. M. and Donzelli, S. and Dor{\'e}, O. and Douspis, M. and Ducout, A. and Dupac, X. and Efstathiou, G. and Elsner, F. and En{\ss}lin, T. A. and Eriksen, H. K. and Falgarone, E. and Fantaye, Y. and Fergusson, J. and {Fernandez-Cobos}, R. and Finelli, F. and Forastieri, F. and Frailis, M. and Franceschi, E. and Frolov, A. and Galeotta, S. and Galli, S. and Ganga, K. and {G{\'e}nova-Santos}, R. T. and Gerbino, M. and Ghosh, T. and {Gonz{\'a}lez-Nuevo}, J. and G{\'o}rski, K. M. and Gratton, S. and Gruppuso, A. and Gudmundsson, J. E. and Hamann, J. and Handley, W. and Hansen, F. K. and Helou, G. and Herranz, D. and Hivon, E. and Huang, Z. and Jaffe, A. H. and Jones, W. C. and Karakci, A. and Keih{\"a}nen, E. and Keskitalo, R. and Kiiveri, K. and Kim, J. and Kisner, T. S. and Knox, L. and Krachmalnicoff, N. and Kunz, M. and {Kurki-Suonio}, H. and Lagache, G. and Lamarre, J. -M. and Langer, M. and Lasenby, A. and Lattanzi, M. and Lawrence, C. R. and Jeune, M. Le and Leahy, J. P. and Lesgourgues, J. and Levrier, F. and Lewis, A. and Liguori, M. and Lilje, P. B. and Lilley, M. and Lindholm, V. and {L{\'o}pez-Caniego}, M. and Lubin, P. M. and Ma, Y. -Z. and {Mac{\'i}as-P{\'e}rez}, J. F. and Maggio, G. and Maino, D. and Mandolesi, N. and Mangilli, A. and {Marcos-Caballero}, A. and Maris, M. and Martin, P. G. and {Mart{\'i}nez-Gonz{\'a}lez}, E. and Matarrese, S. and Mauri, N. and McEwen, J. D. and Meerburg, P. D. and Meinhold, P. R. and Melchiorri, A. and Mennella, A. and Migliaccio, M. and Millea, M. and Mitra, S. and {Miville-Desch{\^e}nes}, M. -A. and Molinari, D. and Moneti, A. and Montier, L. and Morgante, G. and Moss, A. and Mottet, S. and M{\"u}nchmeyer, M. and Natoli, P. and {N{\o}rgaard-Nielsen}, H. U. and Oxborrow, C. A. and Pagano, L. and Paoletti, D. and Partridge, B. and Patanchon, G. and Pearson, T. J. and Peel, M. and Peiris, H. V. and Perrotta, F. and Pettorino, V. and Piacentini, F. and Polastri, L. and Polenta, G. and Puget, J. -L. and Rachen, J. P. and Reinecke, M. and Remazeilles, M. and Renzi, A. and Rocha, G. and Rosset, C. and Roudier, G. and {Rubi{\~n}o-Mart{\'i}n}, J. A. and {Ruiz-Granados}, B. and Salvati, L. and Sandri, M. and Savelainen, M. and Scott, D. and Shellard, E. P. S. and Shiraishi, M. and Sirignano, C. and Sirri, G. and Spencer, L. D. and Sunyaev, R. and {Suur-Uski}, A. -S. and Tauber, J. A. and Tavagnacco, D. and Tenti, M. and Terenzi, L. and Toffolatti, L. and Tomasi, M. and Trombetti, T. and Valiviita, J. and Van Tent, B. and Vibert, L. and Vielva, P. and Villa, F. and Vittorio, N. and Wandelt, B. D. and Wehus, I. K. and White, M. and White, S. D. M. and Zacchei, A. and Zonca, A.},
 doi = {10.1051/0004-6361/201833880},
 eprint = {1807.06205},
 issn = {0004-6361, 1432-0746},
 journal = {Astron. Astrophys.},
 month = {July},
 pages = {A1},
 shorttitle = {{\emph{Planck}} 2018 Results},
 title = {Planck 2018 Results. I. Overview and the Cosmological Legacy of Planck},
 volume = {641},
 year = {2018}
}

@article{priceSparseBayesianMass2021,
 author = {Price, M A and McEwen, J D and Cai, X and Kitching, T D and Wallis, C G R and {(for the LSST Dark Energy Science Collaboration)}},
 doi = {10.1093/mnras/stab1983},
 issn = {0035-8711, 1365-2966},
 journal = {Mon. Not. R. Astron. Soc.},
 month = {July},
 number = {3},
 pages = {3678--3690},
 publisher = {Oxford University Press (OUP)},
 shorttitle = {Sparse Bayesian Mass Mapping with Uncertainties},
 title = {Sparse Bayesian Mass Mapping with Uncertainties: Hypothesis Testing of Structure},
 volume = {506},
 year = {2021}
}

@article{ramanahWienerFilteringPure2019,
 archiveprefix = {arXiv},
 author = {Ramanah, Doogesh Kodi and Lavaux, Guilhem and Wandelt, Benjamin D.},
 doi = {10.1093/mnras/stz2608},
 eprint = {1906.10704},
 issn = {0035-8711, 1365-2966},
 journal = {Mon. Not. R. Astron. Soc.},
 month = {November},
 number = {1},
 pages = {947--961},
 primaryclass = {astro-ph},
 title = {Wiener Filtering and Pure E/B Decomposition of CMB Maps with Anisotropic Correlated Noise},
 volume = {490},
 year = {2019}
}

@article{remyProbabilisticMassMapping2023,
 archiveprefix = {arXiv},
 author = {Remy, Benjamin and Lanusse, Francois and Jeffrey, Niall and Liu, Jia and Starck, Jean-Luc and Osato, Ken and Schrabback, Tim},
 doi = {10.1051/0004-6361/202243054},
 eprint = {2201.05561},
 issn = {0004-6361, 1432-0746},
 journal = {Astron. Astrophys.},
 month = {April},
 pages = {A51},
 primaryclass = {astro-ph},
 title = {Probabilistic Mass Mapping with Neural Score Estimation},
 volume = {672},
 year = {2023}
}

@article{scaramellaEuclidPreparationEuclid2022,
 archiveprefix = {arXiv},
 author = {Scaramella, R. and Amiaux, J. and Mellier, Y. and Burigana, C. and Carvalho, C. S. and Cuillandre, J.-C. and Silva, A. Da and Derosa, A. and Dinis, J. and Maiorano, E. and Maris, M. and Tereno, I. and Laureijs, R. and Boenke, T. and Buenadicha, G. and Dupac, X. and Venancio, L. M. Gaspar and {G{\'o}mez-{\'A}lvarez}, P. and Hoar, J. and Alvarez, J. Lorenzo and Racca, G. D. and {Saavedra-Criado}, G. and Schwartz, J. and Vavrek, R. and Schirmer, M. and Aussel, H. and Azzollini, R. and Cardone, V. F. and Cropper, M. and Ealet, A. and Garilli, B. and Gillard, W. and Granett, B. R. and Guzzo, L. and Hoekstra, H. and Jahnke, K. and Kitching, T. and Meneghetti, M. and Miller, L. and Nakajima, R. and Niemi, S. M. and Pasian, F. and Percival, W. J. and Sauvage, M. and Scodeggio, M. and Wachter, S. and Zacchei, A. and Aghanim, N. and Amara, A. and Auphan, T. and Auricchio, N. and Awan, S. and Balestra, A. and Bender, R. and Bodendorf, C. and Bonino, D. and Branchini, E. and {Brau-Nogue}, S. and Brescia, M. and Candini, G. P. and Capobianco, V. and Carbone, C. and Carlberg, R. G. and Carretero, J. and Casas, R. and Castander, F. J. and Castellano, M. and Cavuoti, S. and Cimatti, A. and Cledassou, R. and Congedo, G. and Conselice, C. J. and Conversi, L. and Copin, Y. and Corcione, L. and Costille, A. and Courbin, F. and Degaudenzi, H. and Douspis, M. and Dubath, F. and Duncan, C. A. J. and Dusini, S. and Farrens, S. and Ferriol, S. and Fosalba, P. and Fourmanoit, N. and Frailis, M. and Franceschi, E. and Franzetti, P. and Fumana, M. and Gillis, B. and Giocoli, C. and Grazian, A. and Grupp, F. and Haugan, S. V. H. and Holmes, W. and Hormuth, F. and Hudelot, P. and Kermiche, S. and Kiessling, A. and Kilbinger, M. and Kohley, R. and Kubik, B. and K{\"u}mmel, M. and Kunz, M. and {Kurki-Suonio}, H. and Ligori, S. and Lilje, P. B. and Lloro, I. and Mansutti, O. and Marggraf, O. and Markovic, K. and Marulli, F. and Massey, R. and Maurogordato, S. and Melchior, M. and Merlin, E. and Meylan, G. and Mohr, J. J. and Moresco, M. and Morin, B. and Moscardini, L. and Munari, E. and Nichol, R. C. and Padilla, C. and Paltani, S. and Peacock, J. and Pedersen, K. and Pettorino, V. and Pires, S. and Poncet, M. and Popa, L. and Pozzetti, L. and Raison, F. and Rebolo, R. and Rhodes, J. and Rix, H.-W. and Roncarelli, M. and Rossetti, E. and Saglia, R. and Schneider, P. and Schrabback, T. and Secroun, A. and Seidel, G. and Serrano, S. and Sirignano, C. and Sirri, G. and Skottfelt, J. and Stanco, L. and Starck, J. L. and {Tallada-Cresp{\'i}}, P. and Tavagnacco, D. and Taylor, A. N. and Teplitz, H. I. and {Toledo-Moreo}, R. and Torradeflot, F. and Trifoglio, M. and Valentijn, E. A. and Valenziano, L. and Kleijn, G. A. Verdoes and Wang, Y. and Welikala, N. and Weller, J. and Wetzstein, M. and Zamorani, G. and Zoubian, J. and Andreon, S. and Baldi, M. and Bardelli, S. and Boucaud, A. and Camera, S. and Fabbian, G. and Farinelli, R. and {Graci{\'a}-Carpio}, J. and Maino, D. and Medinaceli, E. and Mei, S. and Neissner, C. and Polenta, G. and Renzi, A. and Romelli, E. and Rosset, C. and Sureau, F. and Tenti, M. and Vassallo, T. and Zucca, E. and Baccigalupi, C. and {Balaguera-Antol{\'i}nez}, A. and Battaglia, P. and Biviano, A. and Borgani, S. and Bozzo, E. and Cabanac, R. and Cappi, A. and Casas, S. and Castignani, G. and {Colodro-Conde}, C. and Coupon, J. and Courtois, H. M. and Cuby, J. and de la Torre, S. and Desai, S. and Ferdinando, D. Di and Dole, H. and Fabricius, M. and Farina, M. and Ferreira, P. G. and Finelli, F. and {Flose-Reimberg}, P. and Fotopoulou, S. and Galeotta, S. and Ganga, K. and Gozaliasl, G. and Hook, I. M. and Keihanen, E. and Kirkpatrick, C. C. and Liebing, P. and Lindholm, V. and Mainetti, G. and Martinelli, M. and Martinet, N. and Maturi, M. and McCracken, H. J. and Metcalf, R. B. and Morgante, G. and Nightingale, J. and Nucita, A. and Patrizii, L. and Potter, D. and Riccio, G. and S{\'a}nchez, A. G. and Sapone, D. and Schewtschenko, J. A. and Schultheis, M. and Scottez, V. and Teyssier, R. and Tutusaus, I. and Valiviita, J. and Viel, M. and Vriend, W. and Whittaker, L.},
 doi = {10.1051/0004-6361/202141938},
 eprint = {2108.01201},
 issn = {0004-6361, 1432-0746},
 journal = {Astron. Astrophys.},
 month = {June},
 pages = {A112},
 primaryclass = {astro-ph},
 shorttitle = {Euclid Preparation},
 title = {Euclid Preparation: I. The Euclid Wide Survey},
 volume = {662},
 year = {2022}
}

@article{schrabbackPreciseWeakLensing2018,
 author = {Schrabback, Tim and Schirmer, Mischa and Van Der Burg, Remco F. J. and Hoekstra, Henk and Buddendiek, Axel and Applegate, Douglas and Brada{\v c}, Maru{\v s}a and Eifler, Tim and Erben, Thomas and Gladders, Michael D. and {Hern{\'a}ndez-Mart{\'i}n}, Beatriz and Hildebrandt, Hendrik and Hoag, Austin and Klaes, Dominik and Von Der Linden, Anja and Marchesini, Danilo and Muzzin, Adam and Sharon, Keren and Stefanon, Mauro},
 doi = {10.1051/0004-6361/201731730},
 issn = {0004-6361, 1432-0746},
 journal = {Astron. Astrophys.},
 month = {February},
 pages = {A85},
 shorttitle = {Precise Weak Lensing Constraints from Deep High-Resolution {\emph{K}}{\textsubscript{s}} Images},
 title = {Precise Weak Lensing Constraints from Deep High-Resolution {\emph{K}}{\textsubscript{s}} Images: VLT/HAWK-I Analysis of the Super-Massive Galaxy Cluster RCS2 J 232727.7-020437 at {\emph{z}} = 0.70},
 volume = {610},
 year = {2018}
}

@article{shiAccurateKappaReconstruction2024,
 archiveprefix = {arXiv},
 author = {Shi, Yuan and Zhang, Pengjie and Sun, Zeyang and Wang, Yihe},
 doi = {10.1103/PhysRevD.109.123530},
 eprint = {2311.00316},
 issn = {2470-0010, 2470-0029},
 journal = {Phys. Rev. D},
 month = {June},
 number = {12},
 pages = {123530},
 primaryclass = {astro-ph},
 title = {Accurate Kappa Reconstruction Algorithm for Masked Shear Catalog},
 volume = {109},
 year = {2024}
}

@article{shiAKRA20Accurate2024,
 author = {Shi, Yuan and Zhang, Pengjie and Deng, Furen and Zhou, Shuren and Cai, Hongbo and Yao, Ji and Sun, Zeyang},
 doi = {10.1088/1475-7516/2025/07/038},
 issn = {1475-7516},
 journal = {J. Cosmol. Astropart. Phys.},
 month = {July},
 number = {7},
 pages = {38},
 shorttitle = {Akra 2.0},
 title = {AKRA 2.0: Accurate Kappa Reconstruction Algorithm for Masked Shear Catalog},
 volume = {2025},
 year = {2025}
}

@article{shirasakiNoiseReductionWeak2021,
 archiveprefix = {arXiv},
 author = {Shirasaki, Masato and Moriwaki, Kana and Oogi, Taira and Yoshida, Naoki and Ikeda, Shiro and Nishimichi, Takahiro},
 doi = {10.1093/mnras/stab982},
 eprint = {1911.12890},
 issn = {0035-8711, 1365-2966},
 journal = {Mon. Not. R. Astron. Soc.},
 month = {April},
 number = {2},
 pages = {1825--1839},
 primaryclass = {astro-ph},
 shorttitle = {Noise Reduction for Weak Lensing Mass Mapping},
 title = {Noise Reduction for Weak Lensing Mass Mapping: An Application of Generative Adversarial Networks to Subaru Hyper Suprime-Cam First-Year Data},
 volume = {504},
 year = {2021}
}

@article{shuttleworthLoRAVsFull2024,
 archiveprefix = {arXiv},
 author = {Shuttleworth, Reece and Andreas, Jacob and Torralba, Antonio and Sharma, Pratyusha},
 doi = {10.48550/ARXIV.2410.21228},
 eprint = {2410.21228},
 journal = {Corr},
 primaryclass = {cs},
 shorttitle = {LoRA vs Full Fine-Tuning},
 title = {LoRA vs Full Fine-Tuning: An Illusion of Equivalence},
 volume = {abs/2410.21228},
 year = {2024}
}

@article{starckWeakLensingMass2021,
 archiveprefix = {arXiv},
 author = {Starck, J. -L. and Themelis, K. E. and Jeffrey, N. and Peel, A. and Lanusse, F.},
 doi = {10.1051/0004-6361/202039451},
 eprint = {2102.04127},
 issn = {0004-6361, 1432-0746},
 journal = {Astron. Astrophys.},
 month = {February},
 pages = {A99},
 title = {Weak Lensing Mass Reconstruction Using Sparsity and a Gaussian Random Field},
 volume = {649},
 year = {2021}
}

@article{stephenImprovedNormalizationTimelapse2014,
 author = {Stephen, Karl D. and Kazemi, Alireza},
 doi = {10.1111/1365-2478.12109},
 issn = {0016-8025, 1365-2478},
 journal = {Geophysical Prospecting},
 month = {September},
 number = {5},
 pages = {1009--1027},
 title = {Improved Normalization of Time-lapse Seismic Data Using Normalized Root Mean Square Repeatability Data to Improve Automatic Production and Seismic History Matching in the Nelson Field},
 volume = {62},
 year = {2014}
}

@article{vaswaniAttentionAllYou2023,
 archiveprefix = {arXiv},
 author = {Vaswani, Ashish and Shazeer, Noam and Parmar, Niki and Uszkoreit, Jakob and Jones, Llion and Gomez, Aidan N. and Kaiser, Lukasz and Polosukhin, Illia},
 doi = {10.48550/arXiv.1706.03762},
 eprint = {1706.03762},
 journal = {Neural Inf. Process. Syst.},
 month = {June},
 primaryclass = {cs},
 title = {Attention Is All You Need},
 year = {2017}
}

@article{virtanenSciPy10Fundamental2020,
 author = {Virtanen, Pauli and Gommers, Ralf and Oliphant, Travis E. and Haberland, Matt and Reddy, Tyler and Cournapeau, David and Burovski, Evgeni and Peterson, Pearu and Weckesser, Warren and Bright, Jonathan and Van Der Walt, St{\'e}fan J. and Brett, Matthew and Wilson, Joshua and Millman, K. Jarrod and Mayorov, Nikolay and Nelson, Andrew R. J. and Jones, Eric and Kern, Robert and Larson, Eric and Carey, C J and Polat, {\.I}lhan and Feng, Yu and Moore, Eric W. and VanderPlas, Jake and Laxalde, Denis and Perktold, Josef and Cimrman, Robert and Henriksen, Ian and Quintero, E. A. and Harris, Charles R. and Archibald, Anne M. and Ribeiro, Ant{\^o}nio H. and Pedregosa, Fabian and Van Mulbregt, Paul and {SciPy 1.0 Contributors} and Vijaykumar, Aditya and Bardelli, Alessandro Pietro and Rothberg, Alex and Hilboll, Andreas and Kloeckner, Andreas and Scopatz, Anthony and Lee, Antony and Rokem, Ariel and Woods, C. Nathan and Fulton, Chad and Masson, Charles and H{\"a}ggstr{\"o}m, Christian and Fitzgerald, Clark and Nicholson, David A. and Hagen, David R. and Pasechnik, Dmitrii V. and Olivetti, Emanuele and Martin, Eric and Wieser, Eric and Silva, Fabrice and Lenders, Felix and Wilhelm, Florian and Young, G. and Price, Gavin A. and Ingold, Gert-Ludwig and Allen, Gregory E. and Lee, Gregory R. and Audren, Herv{\'e} and Probst, Irvin and Dietrich, J{\"o}rg P. and Silterra, Jacob and Webber, James T and Slavi{\v c}, Janko and Nothman, Joel and Buchner, Johannes and Kulick, Johannes and Sch{\"o}nberger, Johannes L. and De Miranda Cardoso, Jos{\'e} Vin{\'i}cius and Reimer, Joscha and Harrington, Joseph and Rodr{\'i}guez, Juan Luis Cano and {Nunez-Iglesias}, Juan and Kuczynski, Justin and Tritz, Kevin and Thoma, Martin and Newville, Matthew and K{\"u}mmerer, Matthias and Bolingbroke, Maximilian and Tartre, Michael and Pak, Mikhail and Smith, Nathaniel J. and Nowaczyk, Nikolai and Shebanov, Nikolay and Pavlyk, Oleksandr and Brodtkorb, Per A. and Lee, Perry and McGibbon, Robert T. and Feldbauer, Roman and Lewis, Sam and Tygier, Sam and Sievert, Scott and Vigna, Sebastiano and Peterson, Stefan and More, Surhud and Pudlik, Tadeusz and Oshima, Takuya and Pingel, Thomas J. and Robitaille, Thomas P. and Spura, Thomas and Jones, Thouis R. and Cera, Tim and Leslie, Tim and Zito, Tiziano and Krauss, Tom and Upadhyay, Utkarsh and Halchenko, Yaroslav O. and {V{\'a}zquez-Baeza}, Yoshiki},
 doi = {10.1038/s41592-019-0686-2},
 issn = {1548-7091, 1548-7105},
 journal = {Nat. Methods},
 month = {March},
 number = {3},
 pages = {261--272},
 shorttitle = {SciPy 1.0},
 title = {SciPy 1.0: Fundamental Algorithms for Scientific Computing in Python},
 volume = {17},
 year = {2020}
}

@book{wiener1949extrapolation,
 author = {Wiener, Norbert},
 isbn = {978-0-262-25719-0},
 publisher = {MIT},
 title = {Extrapolation, Interpolation, and Smoothing of Stationary Time Series: With Engineering Applications},
 volume = {7},
 year = {1949}
}

@article{wolfHuggingFacesTransformersStateoftheart2020,
 archiveprefix = {arXiv},
 author = {Wolf, Thomas and Debut, Lysandre and Sanh, Victor and Chaumond, Julien and Delangue, Clement and Moi, Anthony and Cistac, Pierric and Rault, Tim and Louf, R{\'e}mi and Funtowicz, Morgan and Davison, Joe and Shleifer, Sam and {von Platen}, Patrick and Ma, Clara and Jernite, Yacine and Plu, Julien and Xu, Canwen and Scao, Teven Le and Gugger, Sylvain and Drame, Mariama and Lhoest, Quentin and Rush, Alexander M.},
 doi = {10.48550/arXiv.1910.03771},
 eprint = {1910.03771},
 journal = {Corr},
 primaryclass = {cs},
 shorttitle = {HuggingFace's Transformers},
 title = {HuggingFace's Transformers: State-of-the-Art Natural Language Processing},
 volume = {abs/1910.3771},
 year = {2019}
}

@article{yaoCSSTWLPreparation2023,
 author = {Yao, Ji and Shan, Huanyuan and Li, Ran and Xu, Youhua and Fan, Dongwei and Liu, Dezi and Zhang, Pengjie and Yu, Yu and Wei, Chengliang and Hu, Bin and Li, Nan and Fan, Zuhui and Xu, Haojie and Guo, Wuzheng},
 doi = {10.1093/mnras/stad3563},
 issn = {0035-8711, 1365-2966},
 journal = {Mon. Not. R. Astron. Soc.},
 month = {November},
 number = {3},
 pages = {5206--5218},
 shorttitle = {{\emph{CSST}} WL Preparation I},
 title = {{\emph{CSST}} WL Preparation I: Forecast the Impact from Non-Gaussian Covariances and Requirements on Systematics Control},
 volume = {527},
 year = {2023}
}

@article{zaroubiWienerReconstructionLargescale1995,
 archiveprefix = {arXiv},
 author = {Zaroubi, S. and Hoffman, Y. and Fisher, K. B. and Lahav, O.},
 doi = {10.1086/176070},
 eprint = {astro-ph/9410080},
 issn = {0004-637X, 1538-4357},
 journal = {Astrophys. J.},
 month = {August},
 pages = {446},
 title = {Wiener Reconstruction of the Large-Scale Structure},
 volume = {449},
 year = {1995}
}

@article{zoncaHealpyEqualArea2019,
 author = {Zonca, Andrea and Singer, Leo and Lenz, Daniel and Reinecke, Martin and Rosset, Cyrille and Hivon, Eric and Gorski, Krzysztof},
 doi = {10.21105/joss.01298},
 issn = {2475-9066},
 journal = {JOSS},
 month = {March},
 number = {35},
 pages = {1298},
 shorttitle = {Healpy},
 title = {Healpy: Equal Area Pixelization and Spherical Harmonics Transforms for Data on the Sphere in Python},
 volume = {4},
 year = {2019}
}

@online{healformergithub,
    author = "",
    title = "https://github.com/lalalabox/healformers",
    url  = "https://github.com/lalalabox/healformers",
    addendum = ""
}

\end{document}